\documentclass{article}
\usepackage{graphicx}
\usepackage{nicefrac}
\usepackage{adjustbox}
\usepackage{multirow}

\usepackage{epsfig}
\usepackage{amscd}
\usepackage{amsmath}
\usepackage{amsfonts}
\usepackage{amssymb}
\usepackage{multirow}
\newtheorem{theorem}{Theorem}
\newtheorem{proposition}{Proposition}

\newtheorem{corollary}{Corollary}
\newtheorem{example}{Example}
\newtheorem{remark}{Remark}

\begin{document}

\title{Consolidating a Link Centered Neural Connectivity  Framework with Directed Transfer Function Asymptotics}


 \author{Luiz A. Baccal\'{a} \and Daniel Y. Takahashi \and
 Koichi Sameshima}
%\institute{L. A. Baccal\'{a} 
%\at Telecommunications and Control Department, Escola Polit\'{e}cnica, University of S\~{a}o Paulo, \\
%Av. Prof. Luciano Gualberto, trav. 3,\# 158, S\~{a}o Paulo, SP, 05508-900\\
%Brazil\\Tel.: +55-11-3091-5508, Fax: +55-11-3091-5319\\\email{baccala@lcs.poli.usp.br}
%\and  D. Y. Takahashi 
% \at Psychology Department, Neuroscience Institute, Princeton University, Princeton, NJ 08540, USA.\\\email{takahashiyd@gmail.com}
% \and
%K. Sameshima
%\at Department of Radiology and Oncology, Faculdade de Medicina, University of S\~{a}o Paulo, S\~{a}o Paulo, Brazil
%\\\email{ksameshi@usp.br}
%}
%%\authorrunning{Short form of author list} % if too long for running head

% %\institute{F. Author \at
%               first address \\
%               Tel.: +123-45-678910\\
%               Fax: +123-45-678910\\
%               \email{fauthor@example.com}           %  \\
% %             \emph{Present address:} of F. Author  %  if needed
%            \and
%            S. Author \at
%               second address
% }


%\keywords{directed transfer function, Granger causality inference, asymptotic
%theory, connectivity detection}



\maketitle
\begin{abstract}
We present a unified mathematical derivation of the asymptotic behaviour of three of the main forms of \textit{directed transfer function} (DTF) complementing recent  partial directed coherence (PDC) results  \cite{Baccala2013}. Based on these results and numerical examples we argue for a new directed `link' centered neural connectivity  framework to replace the widespread correlation based effective/functional network concepts so that directed network influences between structures become classified as to whether links are \textit{active}  in a \textit{direct} or in an \textit{indirect} way thereby leading to the new notions of \textit{Granger connectivity} and \textit{Granger influenciability} which are more descriptive than speaking of Granger causality alone.
\end{abstract}

%\keywords{DTF asymptotics, partial directed coherence, link  centered connectivity framework, Granger connectivity, Granger influenciability}
%\begin{document}

\section{Introduction}
\label{sec:intro}

Introduced as a frequency domain characterization of the interaction between multiple neural structures \textit{directed transfer function} (DTF) \cite{Kaminski91} can be thought as a factor in the coherence between pairs of observed time series \cite{Baccalabiocib2001}. A historical perspective on DTF by their authors can be found in \cite{CRCbBlina} together with its many variants. 

On a par with it, stands \textit{partial directed coherence} (PDC) \cite{Baccalabiocib2001} as its dual measure. The chief distinction between them is that  PDC captures \textit{active} immediate directional coupling between structures whereas DTF, in general, portrays the existence of directional signal propagation even if it is only indirect, by going through intermediate structures rather through immediate direct causal influence \cite{dPDC}. DTF, therefore, represents signal `\textit{reachability}' in a  graph theoretical sense whereas PDC is akin to an adjacency matrix description \cite{harary1994graph}. 

Since DTF's introduction, we examined two of its closely related variants (a) directed coherence (DC) \cite{asp98} which is DTF's scale invariant form (and dual to generalized PDC ($g$PDC)\cite{gPDC2007}) and (b) information DTF ($\iota \text{DTF}$) which is an information theoretic generalization of DTF, dual to information PDC ($\iota$PDC), both of which provide accurate size effect information \cite{Daniel_BA_2010,Takahashi2010,InfoTheo}.

In this paper, we derive and illustrate inference results for the above DTF variants from a unified perspective closely paralleling the  inference results in \cite{Baccala2013} and further illustrated in \cite{CRCgAsympPdc} for PDC and its variants. The importance of accurate asymptotics for DTF is that jointly DTF and PDC allow extending the current paradigm of effective/function connectivity to a more general and informative context \cite{CRCpEpilogue}.  

After briefly reviewing DTF's formulations (Sec.
\ref{sec:prelim}) together with a summary of the unified asymptotic results (Sec. \ref{sec:math_results}), numerical illustrations 
(Sec. \ref{sec:illustr}) discuss some implications of the results as further elaborated in  Sec. \ref{sec:discuss} with their implications  for the new connectivity analysis paradigm we proposed in \cite{CRCpEpilogue}. For reader convenience, mathematical details are left to the Appendix whose implementation is to appear in the next release of the AsympPDC package \cite{Baccala2013}. 

\section{Background}
\label{sec:prelim}

The departure point for defining all DTF related variants is an adequately fitted multivariate autoregressive time series (i.e. vector time series) model to which a multivariate signal $\mathbf{x}(n)$ made up by $x_k(n), \; k=1,\dots, K$ simultaneously acquired time series conforms to
\begin{equation}
\label{eq:model_law}
\mathbf{x}(n)=\sum_{l=1}^{p}\mathbf{A}(l)\mathbf{x}(n-l)+\mathbf{w}(n),
\end{equation}
where $\mathbf{w}(n)$ stands for a zero mean white innovations process of with
$\boldsymbol{\Sigma}_\mathbf{w}=[\sigma_{ij}]$ as its covariance
matrix and $p$ is the model order. The  $a_{ij}(l)$ coefficients composing each $\mathbf{A}(l)$ matrix describe the lagged effect of the $j$-th on the $i$-th series, wherefrom one can also define a frequency domain representation of \eqref{eq:model_law} via the $\bar{\mathbf{A}}(f)$ matrix whose entries are given by
\begin{equation}
\label{acoef}
\bar{A}_{ij}(\lambda)=\left\{ 
\begin{array}{l}
1-\sum\limits_{l=1}^{p}a_{ij}(l)e^{-\mathbf{j}2\pi \lambda l},\;\text{if}\;\;i=j \\ 
-\sum\limits_{l=1}^{p}a_{ij}(l)e^{-\mathbf{j}2\pi \lambda l},\;\text{otherwise}
\end{array}
\right.  
\end{equation}
where $\mathbf{j}=\sqrt{-1}$, so that one may define
\begin{equation}
 \label{eq:dtf_matrix}
\mathbf{H}(\lambda)=\bar{\mathbf{A}}^{-1}(\lambda)
\end{equation}
with $H_{ij}(\lambda)$ entries and rows 
denoted ${\bf h}_i$. This leads to a general expression 
\begin{equation}
\label{eq:full_DTF}
\gamma_{ij}(\lambda)= 
\dfrac{s\;\bar{H}_{ij}(\lambda)}{\sqrt{{\bf h}_i^H(\lambda)%
\mathbf{S}{\bf {h}}_i(\lambda)}}  
\end{equation}
summarizing all the forms of DTF form $j$ to $i$ considered herein. The
superscript $^H$ denotes the usual Hermitian transpose. 
The reader should be forewarned to use the 
adequate expression for $s$ and $\mathbf{S}$ to obtain each DTF variant in
\eqref{eq:full_DTF} using Table \ref{tab:Correspond}.
\begin{center}
\begin{table}[h]
\centering
\caption{Defining variables according to DTF type in ~\eqref{eq:full_DTF}} 
\label{tab:Correspond}       % Give a unique label
\begin{tabular}{cccc}
\hline\noalign{\smallskip}
variable & DTF & DC & $\iota$DTF  \\
\noalign{\smallskip}\hline\noalign{\smallskip}
$s$ & 1 & $\sigma^{\nicefrac{1}{2}}_{ii}$ & $\sigma^{\nicefrac{1}{2}}_{ii}$ \\
\\
$\mathbf{S}$ & $\mathbf{I}_K$ & $(\mathbf{I}_K\odot
\boldsymbol{\Sigma_\mathbf{w}})$ & $\boldsymbol{\Sigma}_\mathbf{w}$ \\
%$\boldsymbol{\mathcal{S}}_n$ &  $\mathbf{\mathcal{I}}_{2K}\otimes\mathbf{\mathcal{I}}_K$ & 
%$\mathbf{\mathcal{I}}_{2K}\otimes(\mathbf{\mathcal{I}}_K\odot\boldsymbol{\Sigma})^{-1} $ & 
%$\mathbf{\mathcal{I}}_{2K}\otimes(\mathbf{\mathcal{I}}_K\odot\boldsymbol{\Sigma})^{-1} $ \\
%$\boldsymbol{\mathcal{S}}_d$ & $\mathbf{\mathcal{I}}_{2K}\otimes\mathbf{\mathcal{I}}_K$ & $\mathbf{\mathcal{I}}_{2K}\otimes(\mathbf{\mathcal{I}}_K\odot\boldsymbol{\Sigma})^{-1} $  & 
%$\mathbf{\mathcal{I}}_{2K}\otimes\boldsymbol{\Sigma}^{-1} $  \\
\noalign{\smallskip}\hline
\end{tabular}
\end{table}
\end{center}

% \subsection{Problem Formulation}
% 
% Along the same lines adopted in \cite{Takahashi2007} for the original PDC, 
% it is convenient to reparametrize \eqref{eq:full_DTF} before
% applying the delta method \cite{Vaart1998} which consists of an appropriate
% Taylor expansion of the distribution
% of \eqref{eq:full_DTF} with respect to the quantities $a_{ij}(r)$ and
% $\boldsymbol{\Sigma}_\mathbf{w}$ that need to be estimated when fitting
% \eqref{eq:model_law} and whose dispersion depends on $n_s$ the number of
% available observations.

%%%=>1*
% To rewrite  \eqref{eq:full_DTF} as a ratio of quadratic forms, 
% \begin{equation}
% \label{eq:pdc_ratio}
%|\gamma_{ij}(\lambda)|^2=\gamma(\boldsymbol{\theta})=\frac{\gamma_n(\boldsymbol
%{\theta } ) } {\gamma_d(\boldsymbol{\theta})}
% \end{equation}
% for the parameter vector $\boldsymbol{\theta}=\left[\mathbf{h}^T
% \boldsymbol{\sigma}^T\right]^T$
% where
% \begin{equation} 
% \label{eq:h_building}
% \mathbf{h}(\lambda)= \begin{bmatrix}
% \mathit{vec}(Re(\mathbf{H}(\lambda))) \\
% \mathit{vec}(Im({\mathbf{H}}(\lambda)))
% \end{bmatrix} 
% \end{equation}
% and
% \begin{equation}
% \boldsymbol{\sigma}=vec\;\boldsymbol{\Sigma}_\mathbf{w}
% \end{equation}
% omitting $i$,$j$ and $\lambda$ dependence to simplify notation.

\section{Result Overview}
\label{sec:math_results}

The statistical behaviour of  \eqref{eq:full_DTF} in terms of the number of times series data points ($n_s$) can be approximated invoking the \textit{delta} method \cite{Vaart1998} consisting of an appropriate Taylor expansion of the statistics, leading, under mild assumptions, to the following results:

\subsection{Confidence Intervals}
\label{sec:Confidence}

In most applications, because $n_s$ is large, usually  only the first Taylor derivative suffices.
In the present context, parameter asymptotic normality implies that DTF's point estimate will also be asymptotically normal, i.e.
\begin{equation}
\label{eq:confidence}
\sqrt{n_s}
(\left|\widehat{\gamma}_{ij}(\lambda)\right|^2-\left|\gamma_{ij}
(\lambda)\right|^2)\stackrel{d}{\rightarrow} \mathcal{N}(0,\gamma^2(\lambda)),
\end{equation}
where $\gamma^2(\lambda)$ is a frequency dependent variance whose full disclosure requires the introduction of further notation
and is postponed to the Appendix.

\subsection{Null Hypothesis Test}
\label{sec:null_hyp}

Under the null hypothesis,
\begin{equation}
\label{eq:null_hyp_1}
H_0:\;\left|{\gamma}_{ij}(\lambda)\right|^2=0
\end{equation}
$\gamma^2(\lambda)$ vanishes identically so that \eqref{eq:confidence} no longer
applies and the next Taylor term becomes necessary \cite{Serfling1980}. 
The next expansion term is quadratic in the parameter vector and corresponds to
one half of DTF's Hessian at the point of interest with an $O(n_s^{-1})$ dependence.

The resulting distribution is that of a sum of at most two properly weighted independent 
$\chi^2_1$  variables where the weights depend on frequency. Explicit computation is left
to the Appendix given the need of specialized notation, but can be summarized as

\begin{equation}
\label{eq:null_distrib_overview}
n_s \; (\mathbf{h}_i^H(\lambda)
\mathbf{S}\mathbf{h}_i(\lambda))(\left|\widehat{\gamma}_{ij}
(\lambda)\right|^2-\left|\gamma_{ij}(\lambda)\right|^2)
\stackrel{d}{\rightarrow} \sum_{k=1}^{q} l_k(\lambda) \chi^2_1
\end{equation}
where $l_k(\lambda)$ are at  most two frequency dependent eigenvalues ($q \leq
2$) coming from a matrix that depends on DTF's values. 
Note that implicit dependence also comes from the left side of
\eqref{eq:null_distrib_overview} on DTF's denominator. See further details in
Proposition \ref{prop:normal}. Brief comments and explicit computational
methods relating sums of $\chi^2_1$  variables may be found in
\cite{Patnaik49}, \cite{Imhof61} and \cite{Takahashi2007}.

These results closely parallel PDC ones in \cite{Baccala2013}, the main difference lying in how the frequency dependent covariance of the parameter vectors are computed. 

\section{Numerical Illustrations}
\label{sec:illustr}
In the examples that follow dashed lines indicate threshold values and gray shades stand for point confidence intervals around significant points. Unless stated otherwise, innovations noise is zero mean, unit variance and mutually uncorrelated.  The frequency domain graphs are displayed in the standard form of an array where grayed diagonal panels contain the estimated time series power spectra.

\begin{figure}[htbp!]
\centering
\includegraphics[width=0.9\linewidth]{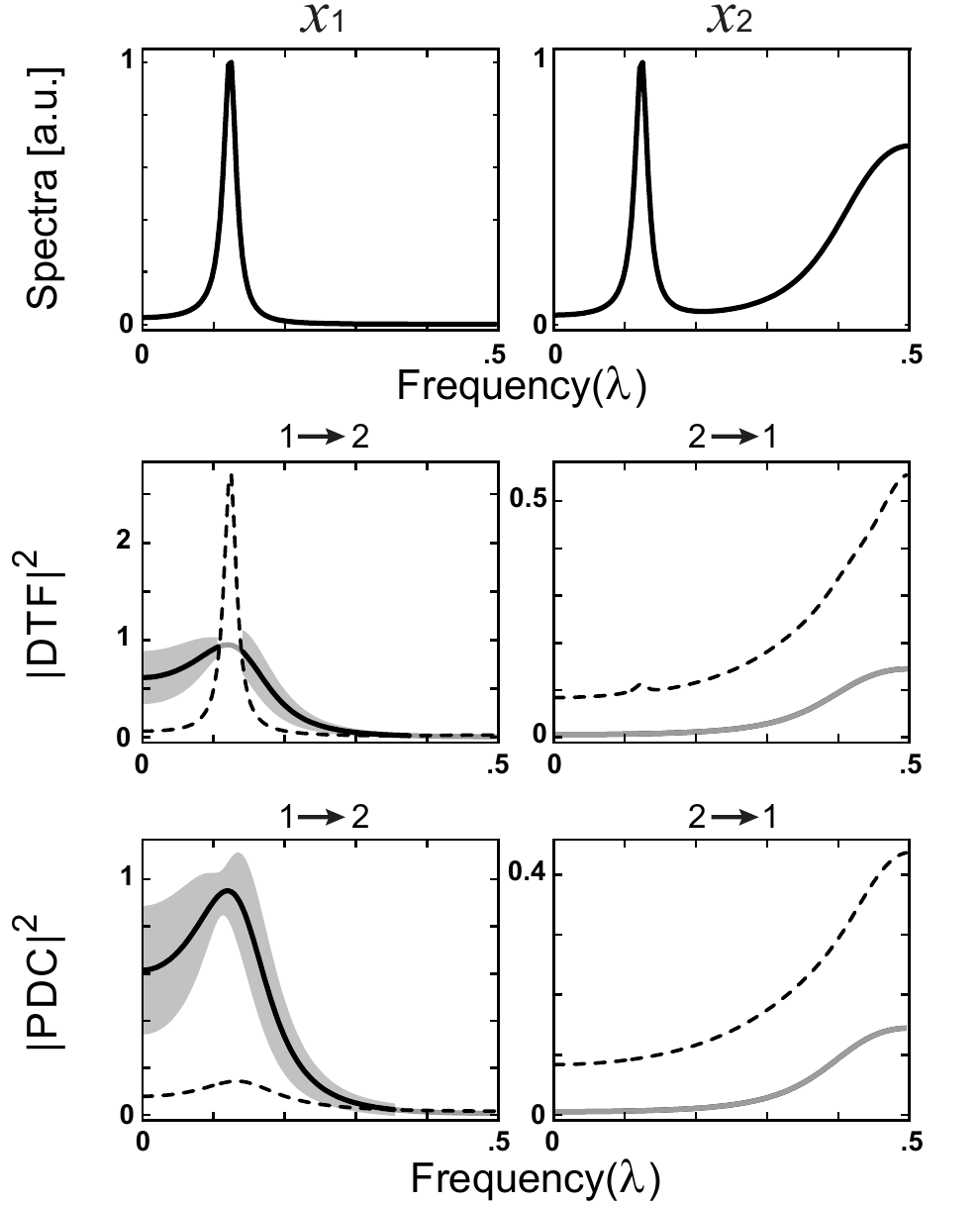}
	\caption{Comparison between DTF (middle row) and PDC (bottom row) for Ex. \protect{\ref{ex:short_resonant_source}} showing the effect of the existing resonance (time series spectra top row) on threshold decision levels (dashed curves) using $n_s=50$ simulated data points. The effect of increasing $n_s$ can be appreciated in Fig.  \protect{\ref{fig:resonant_source}}.
	 Gray shades describe $95\%$ confidence levels when above threshold.} 
	\label{fig:short_resonant_source}
\end{figure}

\begin{figure}[htbp!]
\centering
\includegraphics[width=0.875\linewidth]{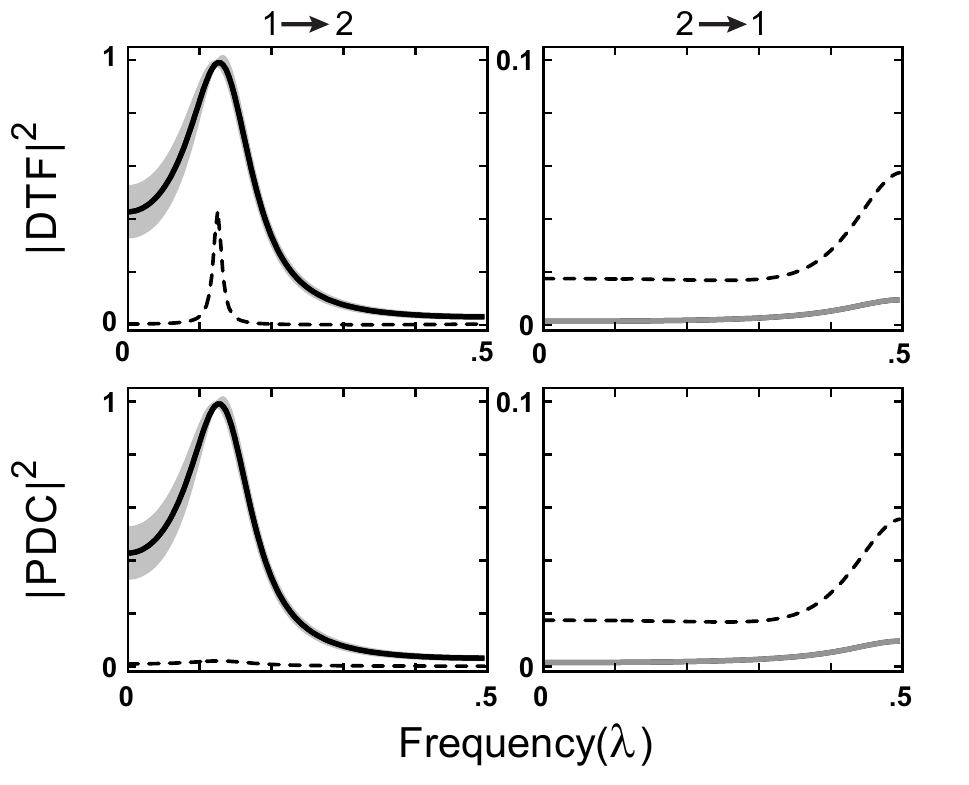}
	\caption{For $n_s=500$ a DTF single trial realization is safely above threshold - compare it to using $n_s=50$ in Fig. \protect{\ref{fig:short_resonant_source}} (middle panel row). Gray shades describe $95\%$ confidence levels when above threshold.} 
	\label{fig:resonant_source}
\end{figure}

\begin{figure}[htbp!]
\centering
\includegraphics[width=\linewidth]{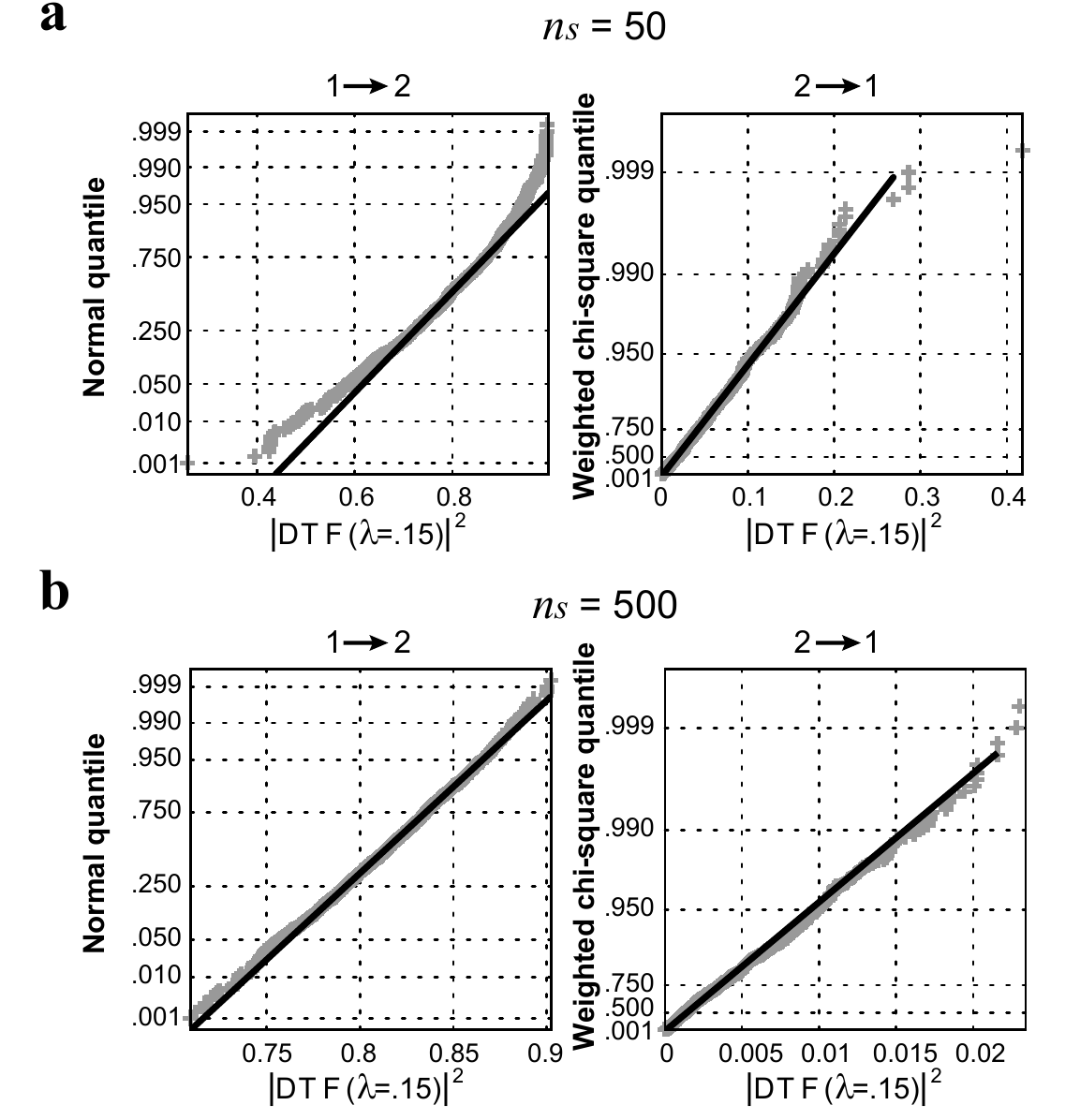}
	\caption{Quantile distribution behaviour showing the distribution goodness of fit improvement as with sample size for Ex. \ref{ex:short_resonant_source} ($n_s=50$ (\textbf{a}) versus $n_s=500$ (\textbf{b})). Statistical spread decrease in the non-existing link $x_2\rightarrow x_1$ is evident as the improved normal fit  of the $x_1\rightarrow x_2$ existing connection. For each value of $n_s$, $m=2000$ simulations were performed.} 
	\label{fig:resonant_source_ensemble}
\end{figure}

\begin{example}
\label{ex:short_resonant_source}
Consider the connectivity from $x_1\rightarrow x_2$ who\-se dynamics is  represented by an oscillator which influences another   structure without feedback:
\begin{eqnarray}
\label{eq:short_resonant_source}
x_{1}(n) &= & 0.95\sqrt{2}x_{1}(n-1) - 0.9025x_{1}(n-2) + w_1(n)\nonumber\\
x_{2}(n)& = & - 0.5x_{1}(n-1) + 0.5x_2(n-1) + w_{2}(n)
\end{eqnarray}

As in all bivariate cases, DTF and PDC coincide numerically, yet because DTF computation requires actual matrix inversion in the general case, its  null hypothesis  threshold limits are affected by the spectra (top panel in Fig. \ref{fig:short_resonant_source})  (the $x_1(n)$ in this case) casting
decision doubts at  $n_s=50$ points (mid panel) as opposed to the PDC case (bottom panel). 
 
 At $n_s=500$, DTF is above threshold  for  $x_1 \rightarrow x_2$  throughout the frequency interval (Fig. \ref{fig:resonant_source}). Further comparison is provided in Fig. \ref{fig:resonant_source_ensemble}\textbf{a} where the actually observed DTF values for $n_s=50$ are more spread than those in Fig. \ref{fig:resonant_source_ensemble}\textbf{b} for $n_s=500$. In the $x_2\rightarrow x_1$ direction, \eqref{eq:null_distrib_overview} behaviour is readily confirmed. 
 
Even though  bivariate DTF and PDC numerically coincide, the need to take into account $\mathbf{\bar{A}}(\lambda)$ inversion under the DTF's null hypothesis can lead to overly conservative thresholds and consequent failure to  properly reject $H_0$ if $n_s$ is small as in Fig. \ref{fig:short_resonant_source}.
\end{example}

\begin{figure}[htb!]
\centering
\includegraphics[width = 0.32\linewidth]{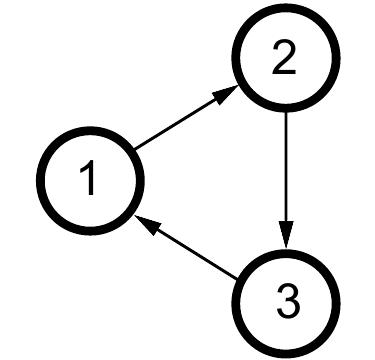}
	\caption{Ex \ref{ex:trivariate_loop} loop connectivity structure. Signals from any structure reach all other structures.} 
	\label{fig:feedback3}
\end{figure}

\begin{figure}[htbp!]
\centering
\includegraphics[width=0.9\linewidth]{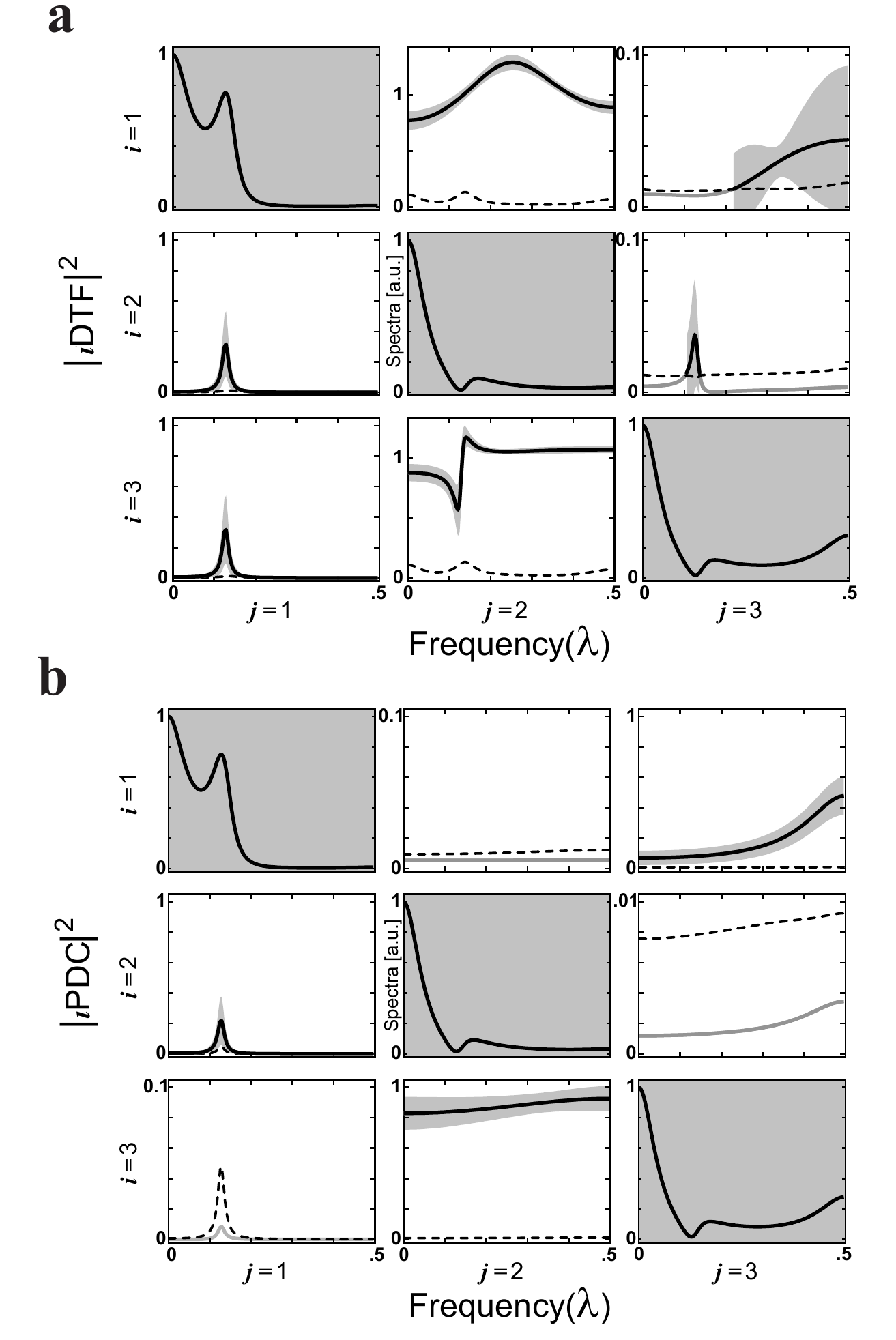}
	\caption{Ex. \ref{ex:trivariate_loop} information $\iota$DTF more widely spread results (\textbf{a}) contrasted to $\iota$PDC (\textbf{b}) results for $n_s=500$ and $\alpha=0.05$. Time series spectra are displayed along the main panel diagonal (gray backgrounds). Sources are marked $j$ (columns) and targets $i$ (rows).} 
	\label{fig:loop3_500}
\end{figure}

\begin{figure}[htbp!]
\centering
\includegraphics[width=0.9\linewidth]{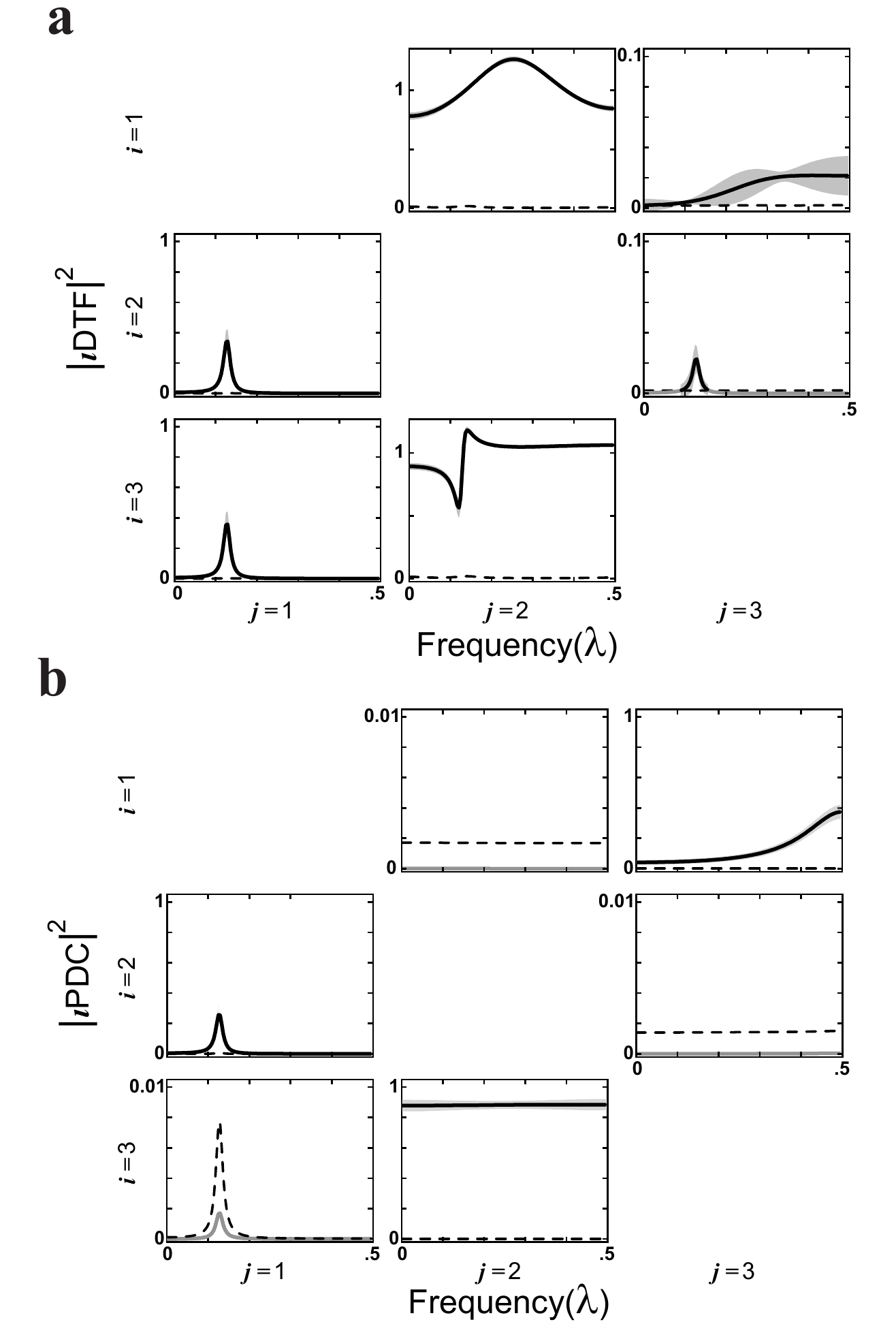}
	\caption{Improvement  of connectivity estimates under $n_s=2000$ over Fig. \ref{fig:loop3_500} single trial behaviour using $\iota$DTF (\textbf{a}) and $\iota$PDC (\textbf{b}). Time series spectra are omitted but can be appreciated from Fig. \protect{\ref{fig:loop3_500}}. Sources are marked $j$ and targets $i$.}
	\label{fig:loop3_2000}
\end{figure}

\begin{figure}[htbp!]
\centering
\includegraphics[width=\linewidth]{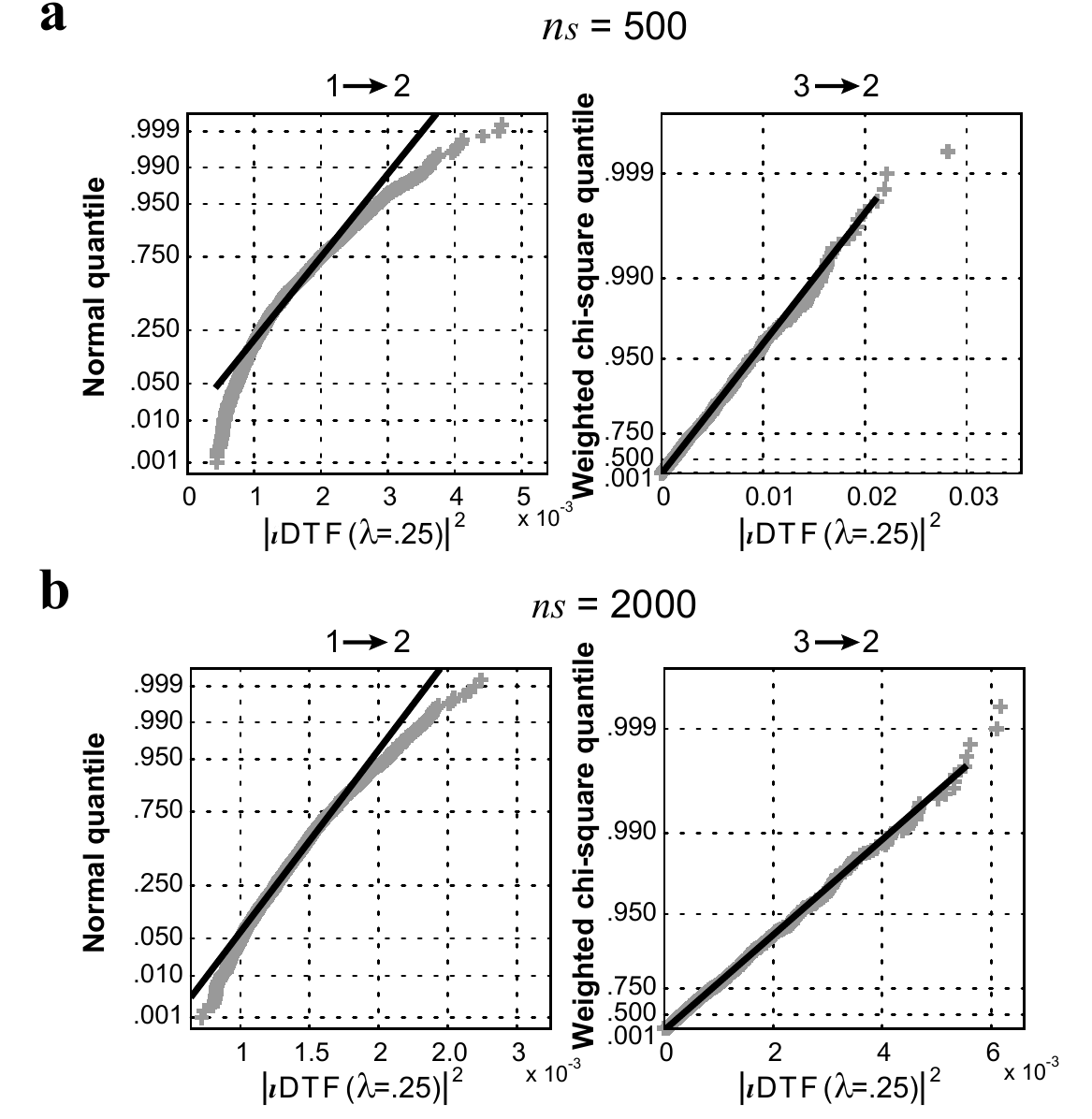}
	\caption{Ex. \ref{ex:trivariate_loop} $\iota$DTF quantile behaviour at $\lambda=0.25$ for $m=2000$ simulations using $n_s=500$ (\textbf{a}) and $n_s=2000$ (\textbf{b}) data points.}
	\label{fig:loop3_qqplots}
\end{figure}

\begin{example}
\label{ex:trivariate_loop}

This example shows an oscillator $x_1(n)$ whose  influence travels back to itself through a loop containing the  $x_2(n)\rightarrow x_3(n)$ link in the feedback loop  pathway (Fig. \ref{fig:feedback3}) and whose dynamics follows:
\begin{eqnarray}
\label{eq:positive_feedback1}
%\begin{array}
 x_1(n) &=&0.95\sqrt{2} x_1(n-1) - 0.9025 x_1(n-2) \nonumber\\
 & & + 0.35 x_3(n-1)+
w_1(n) \nonumber \\
   x_2(n)& =& 0.5 x_1(n-1)+0.5 x_2(n-1) +
w_2(n)\nonumber\\
   x_3(n) &=& x_2(n-1) -0.5 x_3(n-1) + w_3(n)
%\end{array}
\end{eqnarray}%\right.
%\end{equation}
with the covariance matrix of $\mathbf{w}$ given by
\begin{equation}
\label{eq:innov}
\boldsymbol{\Sigma}_\mathbf{w}=\left[
\begin{array}{ccc}
1 & 5 & 0.3\\
5 & 100 & 2\\ 
0.3 & 2 & 1
\end{array}
\right] ,
\end{equation}
ensuring that  $x_2(n)$ contributes a large amount of innovation power to the loop.

Because $\iota$DTF  deals well with unbalanced  innovations, it was used with $n_s=500$ (Fig. \ref{fig:loop3_500}) and $n_s=2000$ (Fig. \ref{fig:loop3_2000}) points leading to the following features: (a) the large $|\iota$DTF$_{i2}|^2$ above 1 for some frequencies are due to the large innovations associated with $x_2(n)$ in \eqref{eq:innov}; (b) except for low $\iota$DTF values that require more points for reliable estimation,  calculations confirm that signals originating at any  structure reach all other structures; and (c) because of the much smaller relative power originating from $x_3(n)$, its influence is much harder to detect. The allied $\iota$PDC, also  shown, confirms which immediate links are directionally active even for $n_s=500$.

It is interesting to observe that $|\iota \text{DTF}_{23}|^2$ has a peak around the $x_1$  resonance  frequency which manifests itself because the innovations originating in $x_3$ ($w_3(n)$) are filtered by passing through the resonant filter represented by structure $x_1$ before reaching $x_2$. 
This same type of influence is not so readily apparent (clear only at $n_s=2000$) in $|\iota$DTF$_{13}|^2$  because the power contributed by
$x_3$ is small with respect to that of other sources reaching $x_1$ around that same resonant frequency. 

The sharp jump in $|\iota$DTF$_{32}|^2$ is a byproduct of the fast phase shift that takes place around $x_1$'s resonance as $x_2$'s signal travels through it to reach $x_3$.

A glimpse of the ensemble $\iota$DTF's behaviour can be appreciated in Fig. \ref{fig:loop3_qqplots} showing how difficult it is to detect it if its values are low.

\end{example}

\begin{figure}[ht!]
\centering
\includegraphics[width=0.64\linewidth]{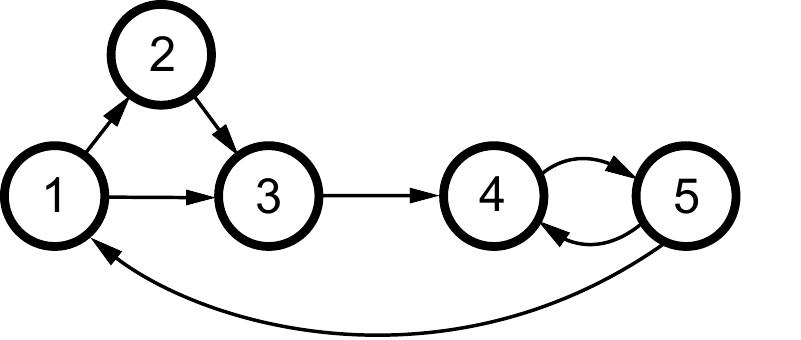}
	\caption{Connectivity diagram for Ex. \ref{ex:closed_loop_big} showing that a signal traveling from any structure can reach any other structure.} 
	\label{fig:closed_loop_big}
\end{figure}

\begin{figure}[htbp!]
\centering
\includegraphics[width=\linewidth]{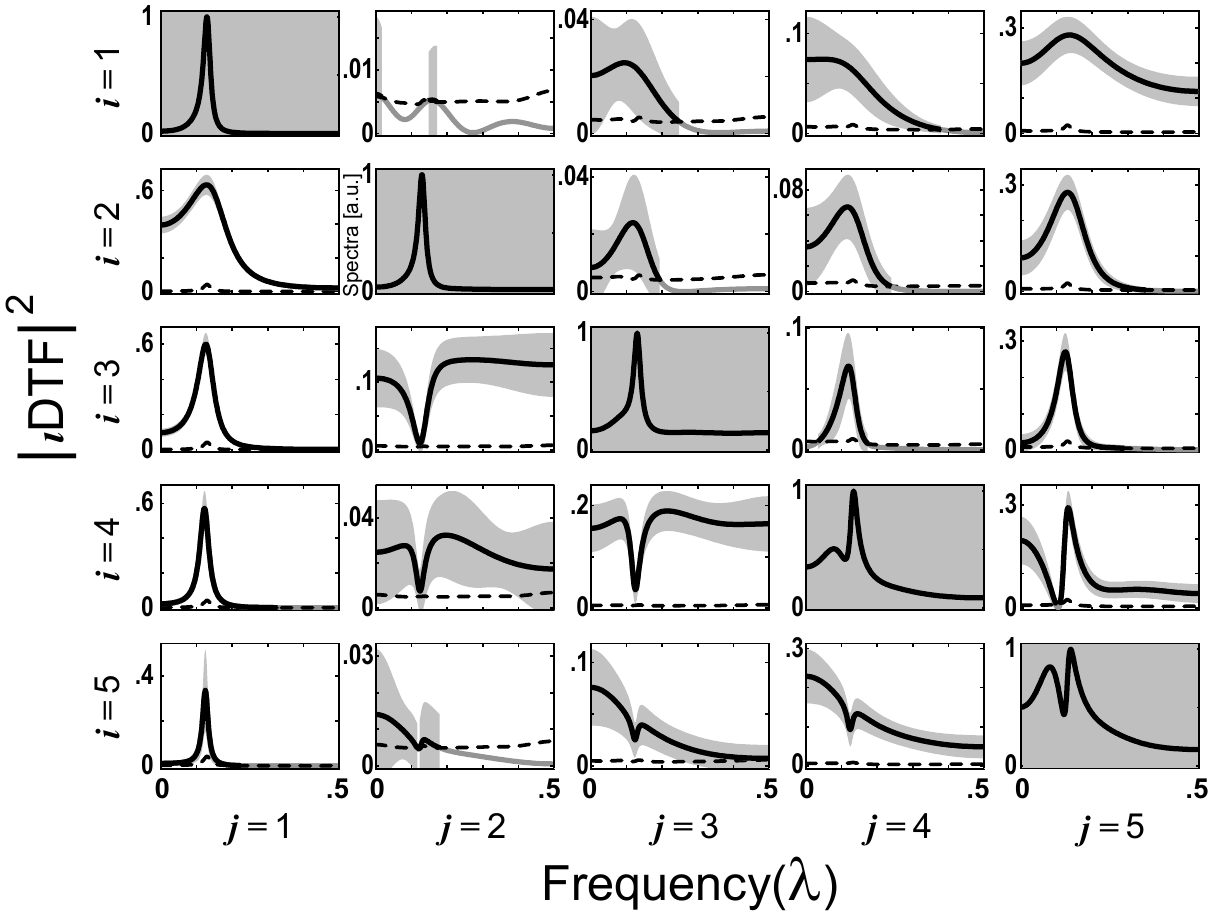}
	\caption{Single  trial $\iota$DTF results for the dynamic connectivity ($n_s=2000$, $\alpha=0.01$) in Ex. \ref{ex:closed_loop_big}. Time series spectra are displayed along the main panel diagonal (gray backgrounds). Note that $\iota$DTF is significant for all pairs (except at some frequencies) confirming that signals from any structures can reach all others. Sources are marked $j$ and targets $i$.} 
	\label{fig:closed_loop_big_dtf}
\end{figure}

\begin{example}
\label{ex:closed_loop_big}

The next example comes from \cite{Baccalabiocib2001}, whose direct connections are contained in Fig. \ref{fig:closed_loop_big} and are dynamically described by:
\begin{eqnarray}
 x_1(n) &=& 0.95\sqrt{2}x_1(n-1) - 0.9025x_1(n-2) \nonumber\\
 & & + 0.5x_5(n-2)+ w_1(n)\nonumber\\
   x_2(n) &=& -0.5x_1(n-1) + w_2(n)\nonumber\\
   x_3(n)&=&0.4x_2(n-2) + w_3(n)\label{eq:baccala_closed_loop_big}\\
   x_4(n)&=&-0.5x_3(n-1) + 0.25\sqrt{2}x_4(n-1)  \nonumber\\ 
   & & + 0.25\sqrt{2}x_5(n-1) + w_4(n)\nonumber\\
   x_5(n)&=&-0.25\sqrt{2}x_4(n-1) + 0.25\sqrt{2}x_5(n-1) + w_5(n)\nonumber
\end{eqnarray}

At first sight, signal pathways exist between all structures, i.e. a signal originating at any structure can reach all other structures.  Note how difficult it is to pinpoint the $x_2$ onto $x_1$ influence even at $n_s=2000$ (Fig. \ref{fig:closed_loop_big_dtf}) throughout the frequency domain interval.

The dynamic  system \eqref{eq:baccala_closed_loop_big} was nonetheless designed to have zero influence at some frequencies as can be appreciated by the dip taken by $|\iota$DTF$_{45}|^2$ whose $x_5\rightarrow x_4$ influence cannot be detected at its minimum even at $n_s=2000$  ($\lambda=0.15$).

\end{example}

\begin{figure}[ht!]
\centering
\includegraphics[width = 0.32\linewidth]{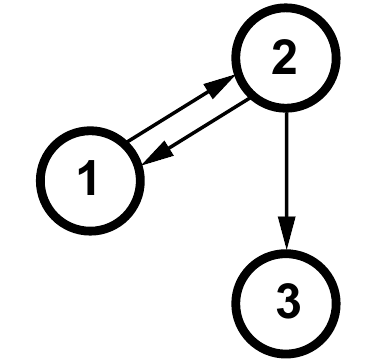}
	\caption{Connectivity diagram between all structures for Ex. \ref{ex:no_dtf_from_3}. Signals from $x_3$ do not reach the other structures.} 
	\label{fig:no_dtf_from_3}
\end{figure}

\begin{figure}[htbp!]
\centering
\includegraphics[width=0.9\linewidth]{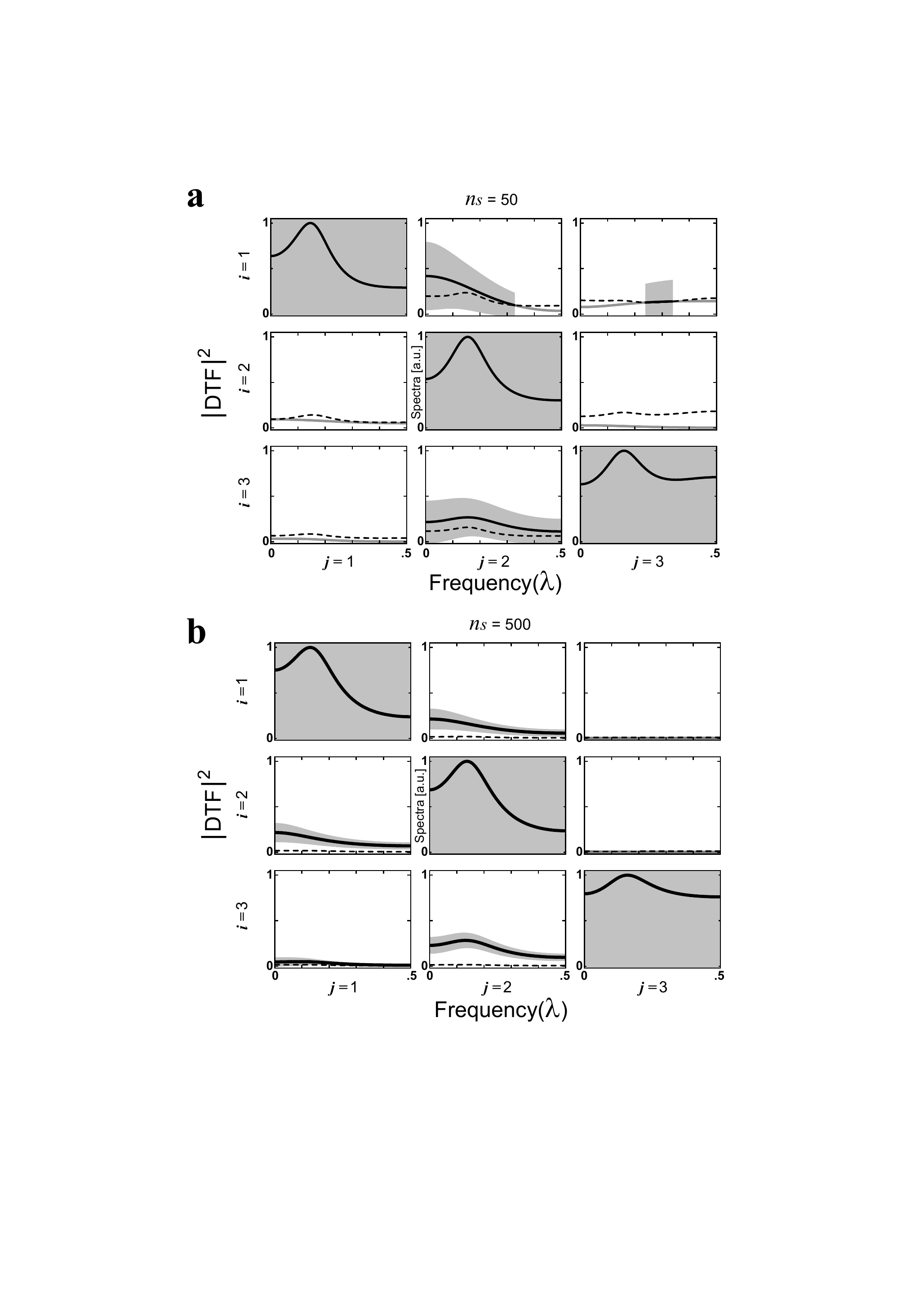}
	\caption{Single trial DTF results for Ex. \ref{ex:no_dtf_from_3} with $n_s=50$ (\textbf{a}) and $n_s=500$ (\textbf{b}), using $\alpha = 0.01$, reflecting the difficulty of inference if $n_s$ is low. Gray shades indicate $99\%$ confidence level for above threshold  $|\text{DTF}|^2$.} 
	\label{fig:no_dtf_from_3_graph}
\end{figure}

\begin{figure}[htbp!]
\centering
\includegraphics[width=\linewidth]{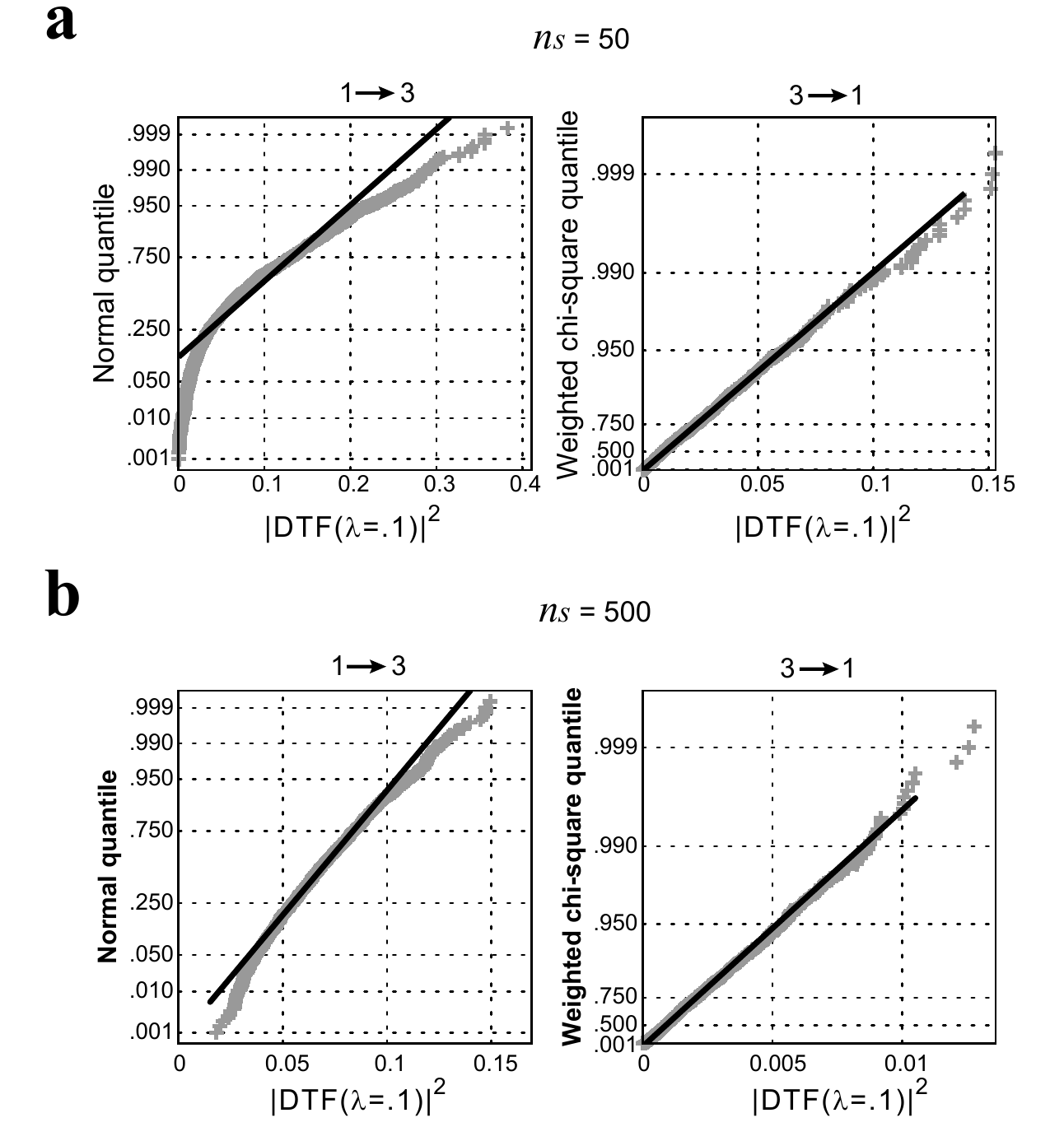}
	\caption{Quantile DTF behaviour  for Ex. \ref{ex:no_dtf_from_3} showing that under $n_s=50$ the asymptotic normality for the existing link $x_1\rightarrow x_3$ is not yet reached (\textbf{a}) as opposed to (\textbf{b}) for $n_s=500$, using $m=2000$ simulations in each case. Reasonable fit for the nonexisting  reverse connection is observed for both $n_s$ values.} 
	\label{fig:no_dtf_from_3_ensemble}
\end{figure}

\begin{example}
\label{ex:no_dtf_from_3}

This last example refers to the connectivity represented in Fig.  \ref{fig:no_dtf_from_3} and behaves dynamically according to

\begin{eqnarray}
 x_1(n)& =&0.95\sqrt{2} x_1(n-1) - 0.9025 x_1(n-2) \nonumber\\
 & & + 0.35 x_2(n-1)+w_1(n)\nonumber \\
   x_2(n)& = &0.5 x_1(n-1)+0.5 x_2(n-1) +w_2(n)\nonumber\\
   x_3(n) &=& x_2(n-1)  -0.5 x_3(n-1) + w_3(n)\label{eq:no_loop}
\end{eqnarray}
whose DTF single trial behaviour can be appreciated in Fig. \ref{fig:no_dtf_from_3_graph}\textbf{a} and \textbf{b} for $n_s=50$ and $n_s=500$ respectively, while Fig. \ref {fig:no_dtf_from_3_ensemble} sums up its ensemble behaviour between $x_1$ and $x_3$.

In this case, there is no $x_3$ influence on other time series, even though detection is more difficult at $n_s=50$ (Fig. \ref{fig:no_dtf_from_3_graph}) highlighting the importance of a sufficiently large sample even for such very simple dynamics. This is confirmed by the point quantile distributions in Fig. \ref{fig:no_dtf_from_3_ensemble}\textbf{a} ($x_1 \rightarrow x_3$) which are somewhat far from the theoretical asymptotic limit, being not yet quite normal compared to the  Fig. \ref{fig:no_dtf_from_3_ensemble}\textbf{b} ($x_1 \rightarrow x_3$) case. 
\end{example}

%The present results are backed up by the software  in the supplementary material that the above examples. A mirror is available at http://www.lcs.poli.usp.br/~baccala/pdc/dftasymp. 

\section{Discussion}
\label{sec:discuss}

The present unified asymptotics for the DTF forms, introduced respectively in \cite{Kaminski91}, \cite{asp98} and \cite{Takahashi2010}, parallel the recent PDC developments in \cite{Baccala2013} and \cite{CRCgAsympPdc}, confirm early  DTF \\ \cite{Kaminski91} threshold results \cite{Eichler2006} and provide estimator confidence intervals and thus pave the way for developing rigorous size effect comparisons using $\iota$DTF and $\iota$PDC under different experimental conditions. 

The allied PDCs presented here should not be confused with the purpose of many recent papers \cite{velez-perez_connectivity_2008, florin_reliability_2011,fasoula_comparative_2013,jovanovic_brain_2013} that compare connectivity techniques, DTF and PDC among them. %Yet an aspect from the examples is that fewer time series points are needed for PDC as specially evident in the bivariate case of Ex. \ref{ex:short_resonant_source} where use of the general DTF estimation matrix inversion introduces  uncertainties that can be avoided given PDC's and DTF's  numerical equality \cite{Baccalabiocib2001,Takahashi2010} making PDC's algorithm the one of choice for time series pairs  when both quantities exactly mirror the  notion of bivariate of Granger causality.

Because many researchers often do not quite realize the conceptual differences between DTF and PDC, this is a good time to stress that they intrinsically measure different aspects of the connectivity between neural structures and neither is conceptually better than the other \cite{jovanovic_brain_2013}, complementary description aspects emerge when  more than just two structures are simultaneously examined. This is clear from Ex. \ref{ex:trivariate_loop} whose loop allows 
signals from any structures to reach all others - as illustrated by DTF's null hypothesis rejection between all structures at least for some frequencies. By contrast PDC shows the overall direction  signals travel within the loop of Ex. \ref{ex:trivariate_loop} by exposing the directional relationship between adjacent structures. 

Furthermore, it is also interesting to realize  (Ex. \ref{ex:closed_loop_big}) that it is entirely possible to have nonzero PDC between pairs of structures and zero DTF between them as happens for DTF from $x_5$ to $x_4$ in this case in the neighbourhood of the network resonant frequency.

Another DTF contribution as to network description is very  well illustrated by the large intrinsic variance of $w_2$ ($x_2$ in Ex. \ref{ex:trivariate_loop}) ensuring  that its influence is comparatively larger upon the other structures; its shape even (see $|\iota$DTF$_{32}|^2$ in Fig. \ref{fig:closed_loop_big_dtf}\textbf{a}) reflects how its originating signal is filtered while traveling through multiple structures. PDC's descriptive interpretations of the direct immediate link between structures can be further examined in \cite{CRCgAsympPdc}.

This newly available objective possibility of performing rigorous hypothesis testing for both DTF and PDC opens up an interesting opportunity in describing neural connectivity. Until now most connectivity has been discussed as either being `effective' and/or `functional', terms coined by \cite{Aertsen1989} and which found widespread employment in functional neuroimaging \cite{Friston1994} becoming  cornerstone descriptive goals for  the recent Connectome effort (see www.connectome.org and  \cite{Perkel}). 

The popularity of  `effective' and `functional' concepts is in line with dynamical network structure descriptions that employ correlation,  an undirected quantity that leads to undirected network graphs \cite{Tononi1994} that stems exactly from the traditional availability of rigorous means for quantifying its statistical reliability even in the light of evidence that its pairwise nature  can lead to difficulties in accurate structural description \cite{Pbrovercoming}.

Furthermore, the effective/functional terminology is confusing:  the `effective' term is reserved for situations when dynamical models of some kind are used for network portrayal whereas `functional' usually refers to perceived statistically significant commonalities (correlations) of dynamical behaviour without special reliance on models. The confusion stems from the possibility of functional connectivity even without effective connectivity as when signal pairs covary under the spell of common  hidden stimuli and no conceivable anatomical links between them.

Through their complementary vistas on connectivity we have proposed a new ``link"  centered framework for network connectivity description \cite{CRCpEpilogue} with PDC describing the immediate direct links between structures  and DTF portraying how one structure ultimately influences another.  In this new framework, directed links (as opposed to undirected ones as provided by correlation) are the central objects of interest. Links now can  be classified  according to their state -- a directed link is in the \textit{direct-active} state when $\text{PDC} \neq 0$, in the \textit{indirect-active} state if $\text{PDC} = 0$ and $\text{DTF} \neq 0$,  is \textit{direct-inactive} for $\text{PDC} =0$ and \textit{indirect-inactive} when $\text{DTF} =0$ as proposed in \cite{Baccala2014} with  the present asymptotics for DTF and PDC \cite{Baccala2013} -- affording the necessary operational state identification tools.

Finally, Granger causality (GC), also popular in neural connectivity descriptions, has had conflicting defenses as to whether DTF or PDC, or even other quantities, represent it in the frequency domain.  Part of the confusion stems from PDC's immediate mirror of the Wald type likelihood ratio tests described as GC tests by \cite{Lutkepohl2005}. Because DTF and PDC indistinguishbly reflect GC in the $K=2$ time series case, but differentiate into distinct network descriptions for  $K>2$, it may worth be developing new terminology to match this differentiation. Within the present link centered framework, we propose to split GC into the complementary views of Granger `connectivity' in association with PDC (graph adjacency) and Granger `influenciability' in connection with `DTF' (graph reachability) both now possessing proper rigorous statistical quantification. Our present introduction of  G-connectivity to describe immediate structural active connections between structures as opposed to possibly long range influences (G-influenciability)  has the intention of standardizing  nomen\-clature and of organizing ideas in our newly proposed paradigm that emphasizes links rover structures as the latter must be significantly active in the first place for the meaningful analysis of their interplay to produce neural function. 

\section*{\protect \centering Appendix}
\appendix
\setcounter{equation}{0}
\renewcommand{\theequation}{A.\arabic{equation}}
%\appendixpage
\section{Propositions}
\label{sec:proofs}

To obtain the results in Sec. \ref{sec:math_results}, specialized notation is needed. The first step is to represent the \eqref{eq:dtf_matrix} as vector of real variables by means of the $vec$ column stacking operator:

\begin{equation}
\label{eq:vec_h}
 \bar{\mathbf{h}}(\lambda)=\mathit{vec} \; \mathbf{H}(\lambda)
\end{equation}
and whose associate real vector can be obtained by noting that
\begin{equation} 
\label{eq:h_building}
\mathbf{h}(\lambda)= \begin{bmatrix}
Re(\bar{\mathbf{h}}(\lambda)) \\
Im(\bar{\mathbf{h}}(\lambda))
\end{bmatrix} 
=\mathbf{\mathcal{T}}
 \begin{bmatrix}
 \bar{\mathbf{h}}(\lambda)\\
 \bar{\mathbf{h}}^*(\lambda)\\
\end{bmatrix}
\end{equation}
where $^*$ denotes complex conjugation and
\begin{equation}
\label{eq:reimag_operator}
 \mathbf{\mathcal{T}}=\dfrac{1}{2}\begin{bmatrix}
                       1 & 1\\
                       -\mathbf{j} & \mathbf{j}
                     \end{bmatrix}
                     \otimes \mathbf{\mathcal{I}}_{K^2}
\end{equation}
for $\mathbf{j}=\sqrt{-1}$.

Also let
\begin{equation}\label{eq:sigma_vec}
\boldsymbol{\sigma}=vec\;\boldsymbol{\Sigma}_\mathbf{w}.
\end{equation}

Together \eqref{eq:vec_h} and \eqref{eq:sigma_vec} can
be combined in the parameter vector
$\boldsymbol{\theta}^T(\lambda)=[\mathbf{h}^T(\lambda)\;
\boldsymbol{\sigma}^T]^T$ so that the problem
of computing DTF, in all its forms, can be reduced to that of computing a ratio
of quadratic forms
\begin{equation}
\label{eq:gPDCmatrix_2}
|\gamma_{ij}(\lambda)|^2=\dfrac{\mathbf{h}^T (\lambda)\mathbf{I}^c_{ij}
\boldsymbol{\mathcal{S}}_n(\boldsymbol{\sigma})\mathbf{I}^c_{ij}\mathbf{h}
(\lambda)}{\mathbf{h}^T (\lambda)
\mathbf{I}^c_{i}\boldsymbol{\mathcal{S}}_d(\boldsymbol{\sigma})\mathbf{I}^c_{i}
\mathbf{h}(\lambda)},
\end{equation}
by employing the definitions
\begin{equation}
\mathbf{I}^c_{ij}=\begin{bmatrix} &\mathbf{I}_{ij} &0\\ &0 &\mathbf{I}_{ij}
\end{bmatrix}
\end{equation}
and
\begin{equation}
\mathbf{I}^c_{i}=\begin{bmatrix} &\mathbf{I}_i &0\\ &0 &\mathbf{I}_i
\end{bmatrix},
\end{equation}
where
\begin{enumerate}
\item $\mathbf{I}_{ij}$ whose entries are zero except for
indices of the form $(l,m)=((j-1)K+i,(j-1)K+i)$, which equal 1 whose purpose is to choose the $i,j$-th element of interest, while
\item $\mathbf{I}_i$ is nonzero only for entries whose indices are of
the form
 $(l,m): l=m=K(r-1)+i$, $1\leq r\leq K$, and  chooses the $i$-th row of interest from the original \eqref{eq:dtf_matrix} matrix,
\end{enumerate}
with $\boldsymbol{\mathcal{S}}_n(\boldsymbol{\sigma})$ and
$\boldsymbol{\mathcal{S}}_d(\boldsymbol{\sigma})$ which do not depend on
$\lambda$ and 
take on different values according to the DTF form under consideration as
listed in Table ~\ref{tab:Correspond}.

The proof of \eqref{eq:gPDCmatrix_2} is immediate and follows of by direct
substitution.

% for each $i$, $j$ and $\lambda$
% 
% 
% The role of \eqref{eq:a_transformation} is to avoid the need of complex
%numbers
% for explicitly describing PDC's magnitude.
% 
% 
% \begin{equation}
% \label{eq:pdc_ratio2}
% \gamma(\boldsymbol{\theta})=\frac{\gamma_n(\boldsymbol{\theta})}{
% \gamma_d(\boldsymbol{\theta})}
% \end{equation} 
% where dependence is left implicit for 

\subsection{Main Asymptotic Results}
The chief basic result is the \textit{delta} method \cite{Vaart1998} which rests on the continuous mapping nature of the parameters onto the statistics of interest in
terms of $n_s$ and the Taylor expansion of the mapping. It is worth summing its
content as: 

\begin{theorem}
\label{theo:gen_delta}
If the distribution of $\mathbf{v}_n=(v_{1}, \,\ldots,\, v_{k})^T$ estimated
from $n$ observations converges in distribution 
as
\begin{equation}
\label{eq:xn_ditrib}
 \sqrt{n}(\mathbf{v}_n-\boldsymbol{\mu})\stackrel{d}{\rightarrow}\mathcal{N}(0,
\mathbf{\Sigma}_\mathbf{v}).
\end{equation}

Let $g(\mathbf{v})$ be a real-valued function with continuous partials of order
$m>1$ in the neighborhood of $\mathbf{v}=\boldsymbol{\mu}$, with all the
partials of order $j$ with $1\leq j\leq m-1$ vanishing at
$\mathbf{v}=\boldsymbol{\mu}$ and non-vanishing $m$-th order partials at
$\mathbf{v}=\boldsymbol{\mu}$. Then
%\begin{equation}
\begin{align}
\label{eq:gendelta} (\sqrt{n})^m
(\hat{g}(\mathbf{v}_n)-g(\boldsymbol{\mu}))\stackrel{d}{\rightarrow}\frac{1}{m!}
\sum_{i_1=1}^k\ldots
  \mspace{180mu}
  \notag\\
\sum_{i_m=1}^k\frac{\partial^mg}{\partial x_{i_1}\ldots
\partial x_{i_m}}\biggr|_{x=\mu_\mathbf{v}}\prod_{j=1}^m\textbf{Z}_{i_j},
\end{align}
%\end{equation}
with
\begin{equation}
\mathbf{Z}=(Z_1, \ldots, Z_k)^T\thicksim
\mathcal{N}(0,\mathbf{\Sigma}_\mathbf{v}), 
\end{equation}
\end{theorem}
\noindent wherefrom one can readily deduce the following consequence for large
$n$ and non
null first derivatives in \eqref{eq:gendelta}:

\begin{corollary}
\label{theo:normal_corollary}
 For a real differentiable function  $g(\mathbf{v})$ asymptotically distributed
as in \eqref{eq:xn_ditrib} then
%$(\lambda))$ (= $|\pi_{ij}(\lambda)|^2$) of the normal vectors $\mathbf{a}(\lambda)$ 
\begin{equation}
\label{eq:normal_corollary}
\sqrt{n}(\hat{g}(\mathbf{v}_{n})-g(\boldsymbol{\mu}))\stackrel{d}{\rightarrow}
\mathcal{N}(0,\mathbf{g}^T\;\boldsymbol{\Sigma}_\mathbf{v}\;\mathbf{g})
\end{equation}
is the first \textit{delta} method approximation where
$\mathbf{g}=\boldsymbol{\nabla}_\mathbf{v} g$ is the gradient of $g(\mathbf{v})$
computed at $\boldsymbol{\mu}$ \cite{Serfling1980}.
\end{corollary}

\begin{remark}
\label{rem:jacobian}
 Though defined for a scalar function of a vector the results from Corollary
\ref{theo:normal_corollary}  remain valid for a vector function where
$\mathbf{g}$ then equals the Jacobian of the transformation rather than the
gradient see p. 26 Sec. 3.1 in \cite{Vaart1998}. 
\end{remark}

\subsection{Asymptotic $\boldsymbol{\theta}$ Behaviour}

\begin{proposition}
\label{lemma:theta_bar_asymp}
The asymptotic properties of $\boldsymbol{\theta}$'s are given by
\begin{equation}
\label{eq:normal_h_bar}
\sqrt{n_s} (
%\hat
\boldsymbol{\hat{\theta}}-\boldsymbol{\theta})\stackrel{d}{\rightarrow}
\mathcal{N}(0,\boldsymbol{\Omega}_{\boldsymbol{\theta}})
\end{equation}
where
\begin{equation}
\boldsymbol{\Omega}_{\boldsymbol{\theta}}=
\begin{bmatrix}
\boldsymbol{\Omega_\mathbf{h}} & \mathbf{0} \\
\mathbf{0} & \boldsymbol{\Omega_{\sigma}}
\end{bmatrix}
\end{equation}
with 
\begin{equation}
\label{h_covariance}
 \boldsymbol{\Omega}_\mathbf{h}(\lambda)=\boldsymbol{\mathcal{H}}(\lambda)
\mathbf{ \mathcal{C}}(\lambda)
\boldsymbol{\Omega}\mathbf{\mathcal{C}}^T(\lambda)\boldsymbol{\mathcal{H}
} ^T(\lambda)
\end{equation}
for
\begin{equation}
\label{eq:hcal}
 \boldsymbol{\mathcal{H}}(\lambda)=-\begin{bmatrix}
                          Re(\mathbf{H}^T(\lambda)\otimes\mathbf{H}(\lambda)) &
-Im(\mathbf{H}^T(\lambda)\otimes\mathbf{H}(\lambda))\\
Im(\mathbf{H}^T(\lambda)\otimes\mathbf{H}(\lambda)) &
Re(\mathbf{H}^T(\lambda)\otimes\mathbf{H}(\lambda))
                          \end{bmatrix}
\end{equation}
and
\begin{equation}
\label{eq:Omega_full}
\boldsymbol{\Omega}=
\begin{bmatrix} 
&\boldsymbol{\Omega}_{\boldsymbol{\alpha}} &
\boldsymbol{\Omega}_{\boldsymbol{\alpha}} \, \\
&\boldsymbol{\Omega}_{\boldsymbol{\alpha}} &
\boldsymbol{\Omega}_{\boldsymbol{\alpha}} \,
\end{bmatrix},
\end{equation}
where $\boldsymbol{\Omega}_{\boldsymbol{\alpha}}$ is the covariance matrix of
\begin{equation}
\label{eq:reorder_coeffs}
\boldsymbol{\alpha}=vec[\mathbf{A}(1)\;\mathbf{A}(2)\;\dots\;\mathbf{A}(p)]
\end{equation}
given by
$\boldsymbol{\Omega_\alpha}=\boldsymbol{\Gamma}_\mathbf{x}^{-1}
\otimes\boldsymbol{\Sigma_w}$, where
$\boldsymbol{\Gamma}_{\mathbf{x}} =E[\mathbf{\bar{x}}(n)
\mathbf{\bar{x}}^{T}(n)]$ for
\begin{align}
\label{eq:x_vector}
%\mathbf{\bar{x}}(n)= [ & x_1(n) \dots x_K(n) x_1(n-1) \dots \nonumber \\
\mathbf{\bar{x}}(n)= [ & x_1(n) \dots x_K(n) \dots \nonumber \\
& x_1(n-p+1) \dots x_K(n-p+1) ]^T;
\end{align}
and
\begin{equation}
\label{eq:omega_sigma}
\boldsymbol{\Omega_{\sigma}}= 2 \mathbf{D_K} \mathbf{D^{+}_K}
(\boldsymbol{\Sigma}_\mathbf{w} \otimes  \boldsymbol{\Sigma}_\mathbf{w})
\mathbf{D}^{+T}_K \mathbf{D}^{T}_K
\end{equation}
with $\mathbf{D}^{+}_K$ standing for the  Moore-Pen\-ro\-se pseudo-inverse of
the
standard duplication matrix \cite{Lutkepohl2005} and
\begin{equation}
\label{eq:cdef}
\mathbf{\mathcal{C}}(\lambda)=
\begin{bmatrix}
\mathbf{C}(\lambda) \\
-\mathbf{S}(\lambda)
\end{bmatrix},
\end{equation}
\topmargin=-.25in
whose blocks are  $K^2 \times pK^2$ dimensional of the form
\begin{equation}
\mathbf{C}(\lambda)=[\mathbf{C}_1(\lambda) \ldots \mathbf{C}_p(\lambda)]
\end{equation}
and
\begin{equation}
\mathbf{S}(\lambda)=[\mathbf{S}_1(\lambda) \ldots \mathbf{S}_p(\lambda)],
\end{equation}
for
\begin{equation}
\mathbf{C}_r(\lambda)=diag([cos(2\pi r \lambda) \ldots cos(2\pi r \lambda)])
\end{equation}
and
\begin{equation}
\mathbf{S}_r(\lambda)=diag([sin(2\pi r \lambda) \ldots sin(2\pi r \lambda)]).
\end{equation}

%and  $\boldsymbol{\Omega_{\sigma_\mathbf{w}}}$ given by \eqref{eq:omega_sigma}.
\end{proposition}
%\begin{proof}
\begin{flushleft}
\textbf{Proof}
\end{flushleft}
 The proof follows by noting that $\mathbf{h}(\lambda)$ is a function of
$\mathbf{a}(\lambda)$ given by
\begin{equation} %\label{a_building}
\label{eq:a_transformation}
\mathbf{a}(\lambda)= \begin{bmatrix}
\mathit{vec}(Re(\bar{\mathbf{A}}(\lambda))) \\
\mathit{vec}(Im(\bar{\mathbf{A}}(\lambda)))
\end{bmatrix} = \begin{bmatrix} vec(\mathbf{\mathcal{I}}_{pK^2})\\
\mathbf{0}\end{bmatrix}-
\mathbf{\mathcal{C}}(\lambda)
\mathbf{\alpha}
\end{equation}
whose covariance matrix is given by \eqref{eq:Omega_full} and is
asymptotically independent from $\boldsymbol{\sigma}$ whose covariance
is \eqref{eq:omega_sigma} as shown in \cite{Baccala2013}.

To obtain $\boldsymbol{\Omega}_\mathbf{h}$, first note that the gradient of
$\bar{\mathbf{h}}$ in \eqref{eq:vec_h} can be obtained from 

\begin{equation}
\nabla_{\bar{\mathbf{a}}} \bar{\mathbf{h}}=-\mathbf{H}^T\otimes\mathbf{H}
\end{equation}
omitting the explicit $\lambda$ dependence to simplify notation,
in view of \eqref{eq:dtf_matrix} in Eq. (1) from
See Sec. 10.6, p. 198 in \cite{HMLutke96}.

By writing 
\begin{equation}
 \begin{bmatrix}
  \nabla_{\bar{\mathbf{a}}} \bar{\mathbf{h}}\\
  \nabla_{\bar{\mathbf{a}}} \bar{\mathbf{h}}^*
 \end{bmatrix}
=
\begin{bmatrix}
 -\mathbf{H}^T\otimes\mathbf{H} & \mathbf{0}\\
  \mathbf{0} & -(\mathbf{H}^T\otimes\mathbf{H})^*
\end{bmatrix}
\begin{bmatrix}
 \bar{\mathbf{a}}\\
 \bar{\mathbf{a}}^*
\end{bmatrix}
\end{equation}
and multiplying by $\mathbf{\mathcal{T}}$ defined in \eqref{eq:h_building} leads
to
\begin{equation}
\nabla_{\mathbf{a}} \mathbf{h}=
\mathbf{\mathcal{T}}
\begin{bmatrix}
 -\mathbf{H}^T\otimes\mathbf{H} & \mathbf{0}\\
  \mathbf{0} & -(\mathbf{H}^T\otimes\mathbf{H})^*
\end{bmatrix}
\mathbf{\mathcal{T}}^{-1}=\mathbf{\mathcal{H}}
\end{equation}

Therefore the result of Remark \ref{rem:jacobian}
implies \eqref{h_covariance} and this completes the proof.
%\end{proof}
\begin{flushright}\ensuremath{\Box}\end{flushright}

\begin{center}
\begin{table}
\centering
\caption{Defining variables according to DTF type in  ~\eqref{eq:gPDCmatrix_2}}
\label{tab:Correspond2}       % Give a unique label
\begin{adjustbox}{width = \linewidth}
\begin{tabular}{cccc}
\hline\noalign{\smallskip}
variable & DTF & DC & $\iota$DTF  \\
\noalign{\smallskip}\hline\noalign{\smallskip}
%$s$ & 1 & $1/\sigma^{1/2}_{ii}$ & $1/\sigma^{1/2}_{ii}$ \\
%$\mathbf{S}$ & $\mathbf{\mathcal{I}}_K$ & $(\mathbf{\mathcal{I}}_K\odot \boldsymbol{\Sigma})^{-1}$ & $\boldsymbol{\Sigma}^{-1}$ \\
$\boldsymbol{\mathcal{S}}_n$ &
$\mathbf{I}_{2K^2}$&
$\mathbf{I}_{2}\otimes(\mathbf{
I}_K\odot\boldsymbol{\Sigma}_\mathbf{w})\otimes \mathbf{I}_{K}$ & 
$\mathbf{I}_{2}\otimes(\mathbf{
I}_K\odot\boldsymbol{\Sigma}_\mathbf{w})\otimes \mathbf{I}_{K}$\\
$\boldsymbol{\mathcal{S}}_d$ &
$\mathbf{I}_{2K^2}$ &
$\mathbf{I}_{2}\otimes(\mathbf{
I}_K\odot\boldsymbol{\Sigma}_\mathbf{w})\otimes \mathbf{I}_{K}$ & 
$\mathbf{I}_{2}\otimes\boldsymbol{\Sigma}_\mathbf{w}\otimes
\mathbf{I}_{K}$  \\
\noalign{\smallskip}\hline
\end{tabular}
\end{adjustbox}
\end{table}
\end{center}
\subsection{Confidence Interval Theorem}
\label{sec:Confidence_proof}

\begin{proposition}
\label{prop:normal}
Omitting the explicit frequency $\lambda$ and $\boldsymbol{\sigma}$ dependencies
to simplify notation, the confidence interval results 
\begin{equation}
\label{eq:confidence2}
\sqrt{n_s}
(\left|\widehat{\gamma}_{ij}\right|^2-\left|\gamma_{ij}\right|^2)\stackrel{d}{
\rightarrow}
\mathcal{N}(0,\gamma^2),
\end{equation}
where $n_s$ is the number of observations and
\begin{equation}
\label{eq:pdc_variance}
\gamma^2=\mathbf{g}_{\mathbf{h}}
\boldsymbol{\Omega}_\mathbf{h}\mathbf{g}^T_{\mathbf{h}}
+
\mathbf{g}_{\boldsymbol{\sigma}}\boldsymbol{\Omega}_{\boldsymbol{\sigma}}\mathbf{g}^T_{\boldsymbol{\sigma}},
\end{equation}
for
\begin{equation}
\label{eq:derivative_alpha}
\mathbf{g}_{\mathbf{h}}=2\frac{\mathbf{h}^T \mathbf{I}^c_{ij}
\boldsymbol{\mathcal{S}}_n\mathbf{I}^c_{ij}}{\mathbf{h}^T \mathbf{I}^c_{i}
\boldsymbol{\mathcal{S}}_d\mathbf{I}^c_{i}\mathbf{h}}
-2\frac{\mathbf{h}^T \mathbf{I}^c_{ij}
\boldsymbol{\mathcal{S}}_n\mathbf{I}^c_{ij}\mathbf{h}}
{(\mathbf{a}^T \mathbf{I}^c_{i}
\boldsymbol{\mathcal{S}}_d\mathbf{I}^c_{i}\mathbf{h})^2}\mathbf{h}^T
\mathbf{I}^c_{i} \boldsymbol{\mathcal{S}}_d\mathbf{I}^c_{i}
\end{equation}
and
\begin{align}
\label{eq:derivative_sigma_theo}
\mathbf{g}_{\boldsymbol{\sigma}} = & 
\frac{1}{\mathbf{h}^T \mathbf{I}^c_{i}
\boldsymbol{\mathcal{S}}_d\mathbf{I}^c_{i}\mathbf{h}}
\left[(\mathbf{I}^c_{ij}\mathbf{h})^T\otimes (\mathbf{h}^T
\mathbf{I}^c_{ij})\right]\boldsymbol{\Theta_K}\boldsymbol{\xi}_n 
 \nonumber \\
& - \frac{\mathbf{h}^T \mathbf{I}^c_{ij}
\boldsymbol{\mathcal{S}}_n\mathbf{I}^c_{ij}\mathbf{h}}{(\mathbf{h}^T
\mathbf{I}^c_{i} \boldsymbol{\mathcal{S}}_d\mathbf{I}^c_{i}\mathbf{h})^2}
\left[(\mathbf{I}^c_{i}\mathbf{h})^T\otimes(\mathbf{h}^T
\mathbf{I}^c_{i})\right]\boldsymbol{\Theta_K}\boldsymbol{\xi}_d
\end{align}
where the values of $\boldsymbol{\xi}_n$ and $\boldsymbol{\xi}_d$ are listed on
Table \ref{tab:xitable} and
\begin{align}\label{eq:Theta_K}
\boldsymbol{\Theta_K}=
(\mathbf{I}_{2}\otimes\mathbf{T}_{K^2,2}
\otimes\mathbf{I_{K^{2}}}) 
  \mspace{230mu}
  \notag\\
\left[ \mathit{vec}
(\mathbf{I}_{2})\otimes
(\mathbf{I}_{K}\otimes\mathbf{T}_{K^2,1}\otimes\mathbf{I}_K)
(\mathbf{I}_{K^2}
\otimes \mathit{vec}(\mathbf{I}_{K})) \right]
\end{align}
%\end{equation}

%\begin{align}
%\label{eq:Theta_K}
%\boldsymbol{\Theta_K}=
%(\mathbf{I}_{2}\otimes\mathbf{T}_{K^2,2}
%\otimes\mathbf{I_{K^{2}}})
%\left[
%\mathit{vec}
%(\mathbf{I}_{2})\otimes
%     \mspace{150mu}
%      \notag\\
%(\mathbf{I}_{K}\otimes\mathbf{T}_{K^2,1}\otimes\mathbf{I}_K) \\
%(\mathbf{I}_{K^2}
%\otimes \mathit{vec}(\mathbf{I}_{K}))\right]
%\end{align}

\noindent with $\mathbf{T}_{L,M}$ standing for the commutation matrix
\cite{Lutkepohl2005}.
\end{proposition}

When the innovation covariance is known \textit{a priori} or does not need to be estimated, the term $\mathbf{g}_{\boldsymbol{\sigma}}\boldsymbol{\Omega}_{\boldsymbol{\sigma}}\mathbf{g}^T_{\boldsymbol{\sigma}}$ in \eqref{eq:pdc_variance} is zero.
\begin{flushleft}
\textbf{Proof}
\end{flushleft}

%\begin{proof}

This proposition is an immediate consequence of Corollary
\ref{theo:normal_corollary} given the results
of Proposition \ref{lemma:theta_bar_asymp} by rewriting \eqref{eq:gPDCmatrix_2}
as

\begin{equation}
\label{eq:gamma_fraction}
\gamma(\boldsymbol{\theta})=
\frac{\gamma_n(\boldsymbol{\theta})}{\gamma_d(\boldsymbol{\theta})}
=\frac{
\mathbf{h}(\lambda)^T \mathbf{I}^c_{ij}
\boldsymbol{\mathcal{S}}_n(\boldsymbol{\sigma})\mathbf{I}^c_{ij}\mathbf{h}
(\lambda)}{\mathbf{h}(\lambda)^T
\mathbf{I}^c_{i}\boldsymbol{\mathcal{S}}_d(\boldsymbol{\sigma})\mathbf{I}^c_{i}
\mathbf{h}(\lambda)}
\end{equation}

All one needs is to properly compute its gradient or that of its transpose.
The job is further simplified by noting the asymptotic independence of
$\mathbf{h}$ and $\boldsymbol{\sigma}$ for
this allows their separate consideration 
as $\boldsymbol{\Omega}_{\boldsymbol{\theta}}$ is block diagonal.
\begin{center}
\begin{table}
\centering
\caption{Defining variables according to DTF type in
\eqref{eq:derivative_sigma_theo}}
\label{tab:xitable}       % Give a unique label
\begin{tabular}{cccc}
\hline\noalign{\smallskip}
variable & DTF & DC & $\iota$DTF  \\
\noalign{\smallskip}\hline\noalign{\smallskip}
$\boldsymbol{\xi}_n $ & 0 &
$\mathit{diag}(\mathit{vec}(\mathbf{I}_{K}))$
&$\mathit{diag}(\mathit{vec}(\mathbf{I}_{K}))$\\
$\boldsymbol{\xi}_d $ & 0 &
$\mathit{diag}(\mathit{vec}(\mathbf{I}_{K}))$
& $\mathbf{I}_{K^2}$ \\
\noalign{\smallskip}\hline
\end{tabular}
\end{table}
\end{center}

Since DTF is a ratio  (see \eqref{eq:gPDCmatrix_2}) one may write the transpose
of the required gradients as
\begin{equation}
\label{eq:divisionrule}
\frac{\partial{\gamma}}{\partial{\boldsymbol{\psi}}^T}=\frac{1}{\gamma_d^2}
\left[\gamma_d \frac{\partial \gamma_n }{\partial{\boldsymbol{\psi}}^T}- 
\gamma_n \frac{\partial \gamma_d }{\partial{\boldsymbol{\psi}}^T}
\right]
\end{equation}
where $\boldsymbol{\psi}$ represents either $\mathbf{h}$ or
$\boldsymbol{\sigma}$ and respectively leads to $\mathbf{g}_{\mathbf{h}}$ and
$\mathbf{g}_{\boldsymbol{\sigma}}$.

Therefore the necessary gradients operate on the required defined quadratic forms whose general
differentiation with respect to $\mathbf{h}$ yields
\begin{equation}
\frac{\partial{(\mathbf{h}^T\mathbf{I}^c
\mathbf{\mathcal{S}}\mathbf{I}^c\mathbf{h})}}{\partial{\boldsymbol{\mathbf{h}}}
^T}=2\mathbf{h}^T\mathbf{I}^c\mathbf{\mathcal{S}}\mathbf{I}^c
\end{equation}
\begin{center}
see p. 175 Eq. (2) in \cite{HMLutke96}
\end{center}
since all appropriately valued $\mathbf{I}^c\mathbf{\mathcal{S}}\mathbf{I}^c$ matrices are symmetric. Inserting
the terms in \eqref{eq:divisionrule} leads to \eqref{eq:derivative_alpha}.

The nested dependence of $\gamma(\boldsymbol{\theta})$ on
$\boldsymbol{\sigma}$ calls for chain rule use:
\begin{equation}
\frac{\partial{(\mathbf{h}^T\mathbf{I}^c\mathbf{\mathcal{S}}\mathbf{I}^c\mathbf{
h})}}{\partial{\boldsymbol{\sigma}}^T}=
\frac{\partial{(\mathbf{h}^T\mathbf{I}^c\mathbf{\mathcal{S}}\mathbf{I}^c\mathbf{
h})}}{\partial{\mathbf{\mathcal{S}}}}
\frac{\partial{\mathbf{\mathcal{S}}}}{\partial{\boldsymbol{\sigma}}^T}
\end{equation}
where 
\begin{equation}
\frac{\partial{(\mathbf{h}^T\mathbf{I}^c\mathbf{\mathcal{S}}\mathbf{I}^c\mathbf{
h})}}{\partial{\mathbf{\mathcal{S}}}}=
(\mathbf{I}^c_{ij}\mathbf{h})^T\otimes (\mathbf{h}^T \mathbf{I}^c_{ij})
\end{equation}
\begin{center}
see p. 183 Eq. (3) in \cite{HMLutke96}
\end{center}
In its general form, $\mathbf{\mathcal{S}}=
\mathbf{I}_{2}\otimes \mathbf{\bar{{\mathcal{S}}}}\otimes
\mathbf{I}_{K}$ ( see Table \ref{tab:Correspond2}). Thus, by the
chain rule

\begin{equation}
\frac{\partial{\mathbf{\mathcal{S}}}}{\partial{\boldsymbol{\sigma}}^T}=
\frac{\partial{\mathbf{\mathcal{S}}}}{\partial{\mathbf{\bar{\mathcal{S}}}}}
\frac{\partial{\mathbf{\bar{\mathcal{S}}}}}{\partial{\boldsymbol{\sigma}}^T}
\end{equation}
where

%\begin{equation}
%\begin{split}
\begin{align}
\frac{\partial{\mathbf{\mathcal{S}}}}{\partial{\mathbf{\bar{\mathcal{S}}}}}=
\boldsymbol{\Theta_K}= 
(\mathbf{I}_{2}\otimes\mathbf{T}_{K^2,2}\otimes\mathbf{I}_{K^{2}}) 
  \mspace{215mu}
  \notag\\
\left[ \mathit{vec}
(\mathbf{I}_{2})\otimes
(\mathbf{I}_{K}\otimes
\mathbf{T}_{K^2,1}\otimes\mathbf{I}_K) (\mathbf{I}_{K^2}
\otimes \mathit{vec}(\mathbf{I}_{K})) \right]
\end{align}
%\end{split}
%\end{equation}

\begin{center}
(see p. 184 Eq. (13) in \cite{HMLutke96}).\\ 
%e tb do ponto de vista de notação a provavel fonte do Carlos - MatCalc.pdf - pg
%8 - dropbox)
\end{center}

All that is left to compute are the derivatives
of $\mathbf{\bar{\mathcal{S}}}$ in each case (as in Table
\ref{tab:Correspond2}). For for each possible $\mathbf{\bar{\mathcal{S}}}$
value, one has
\begin{equation}
\frac{\partial{\mathbf{I}_{K^2}}}{\partial{\boldsymbol{\sigma}}^T}=0
\end{equation}
which follows from its lack of $\boldsymbol{\sigma}$ dependence in this case.

When $\mathbf{\bar{\mathcal{S}}}=\boldsymbol{\Sigma}_\mathbf{w}$,
\begin{equation}
\label{eq:trivial}
\frac{\partial{\boldsymbol{\Sigma}_\mathbf{w}}}{\partial{\boldsymbol{\sigma}}^T}
=\mathbf{I}_{K^2}.
%\boldsymbol{\Sigma}_{\mathbf{w}}^{-1})
\end{equation}

% \begin{equation}
%
%\frac{\partial{\boldsymbol{\Sigma}_\mathbf{w}^{-1}}}{\partial{\boldsymbol{
%\sigma }}^T}=-\boldsymbol{\Sigma}^{-T}_\mathbf{w}\otimes
%\boldsymbol{\Sigma}^{-1}_\mathbf{w}
% \end{equation}
% \begin{center}
% (see \cite{JanR.Magnus11953} p. 183 eq. 19)
% \end{center}
Finally considering
$\mathbf{\tilde{S}}=\mathbf{I}_K\odot\boldsymbol{\Sigma}_\mathbf{w}$
% \begin{equation}
% \label{eq:doublechain}
% \frac{
% \partial{(\mathbf{\tilde{S}})}}{
% {\partial{\boldsymbol{\sigma}}^T}}=
% \frac{\partial{(\mathbf{\tilde{S}}^{-1})}}{\partial{\mathbf{\tilde{S}}}}
% %
% \frac{\partial{\mathbf{\tilde{S}}}}{\partial{\boldsymbol{\Sigma}_\mathbf{w}}}
% \frac{\partial{\boldsymbol{\Sigma}_\mathbf{w}}}{\partial{\boldsymbol{\sigma}}}
% %\boldsymbol{\Sigma}_{\mathbf{w}}^{-1})
% \end{equation}
leads to
 %\begin{equation}
% \frac{\partial{(\mathbf{\tilde{S}})}}{\partial{\mathbf{\tilde{S}}}}=-
% \mathbf{\tilde{S}}^{-T}\otimes
%\mathbf{\tilde{S}}^{-1}=-(\mathbf{\mathcal{I}}_K\odot\boldsymbol{\Sigma}
%_\mathbf {w})^{-T} \otimes
% (\mathbf{\mathcal{I}}_K\odot\boldsymbol{\Sigma}_\mathbf{w})^{-1}
% \end{equation}
%so that usinh
\begin{equation}
\frac{\partial{\mathbf{\tilde{S}}}}{\partial{\boldsymbol{\sigma}^T}}
=diag(vec(\mathbf{I}_K))
%\boldsymbol{\Sigma}_{\mathbf{w}}^{-1})
\end{equation}
\begin{center}
see p. 185 eq. (16) in \cite{HMLutke96}.
\end{center}
% and trivially
% which together in \eqref{eq:doublechain} reduce to 
% 
% \begin{equation}
%
%\frac{\partial{\mathbf{\bar{\mathcal{S}}}}}{\partial{\boldsymbol{\sigma}}}
%=-diag(vec((\mathbf{\mathcal{I}}_K\odot\boldsymbol{\Sigma}_\mathbf{w})^{-2}
%))=-\mathit{diag}(\mathit{vec}(\boldsymbol{\mathcal{S}}_n^{-2}))
% %\boldsymbol{\Sigma}_{\mathbf{w}}^{-1})
% \end{equation}
% \begin{center}
% %(see ***)\\ Better Justification required
% \end{center}
%after some additional algebra. 
These  results are summarized as $\boldsymbol{\xi}_n $ and
$\boldsymbol{\xi}_d $ quantities that appear on Table \ref{tab:xitable} and comprise
\eqref{eq:derivative_sigma_theo}.

The use of Slutsky's lemma concludes the proof by allowing the use of estimated quantities.
%\end{proof}
\begin{flushright}\ensuremath{\Box}\end{flushright} 
\subsection{Null Hypothesis Test}
\label{sec:null_hyp2}

Under the null hypothesis
\begin{equation}
\label{eq:null_hyp}
H_0:\;\left|{\gamma}_{ij}\right|^2=0\;\Longleftrightarrow
\mathbf{I}^c_{ij}\mathbf{h}=\mathbf{0}
\end{equation}
both \eqref{eq:derivative_alpha} and \eqref{eq:derivative_sigma_theo} equal zero, and
\eqref{eq:confidence} no longer applies so that the next Taylor term becomes necessary \cite{Serfling1980} 
weighted by one half of PDC's Hessian at the point of interest with an $O(n_s^{-1})$ dependence.
Via a device similar to that used in \cite{Takahashi2007}, one can show that 

\begin{proposition}
\label{prop:imhof}
Under \eqref{eq:null_hyp}
\begin{equation}
\label{eq:null_distrib}
n_s
(\mathbf{h}\mathbf{I}^c_{i}\boldsymbol{\mathcal{S}}_d\mathbf{I}^c_{i}\mathbf{h})
(\left|\widehat{\gamma}_{ij}\right|^2-\left|\gamma_{ij}\right|^2)
\stackrel{d}{\rightarrow} \sum_{k=1}^{q} l_k \chi^2_1
\end{equation}
where $l_k$ are the eigenvalues of
$\mathbf{\mathcal{D}}=\mathbf{L}^T\mathbf{I}^c_{ij}\boldsymbol{\mathcal{S}}_n \mathbf{I}^c_{ij}\mathbf{L}$, where $\mathbf{L}$ is the Choleski factor of
$\boldsymbol{\Omega}_\mathbf{h}$. Furthermore
$q=rank(\mathbf{\mathcal{D}})\leq2$, its value is $1$
whenever $\lambda \in \left\lbrace  0, \pm 0.5 \right\rbrace $. 
\end{proposition}

The result in \eqref{eq:null_distrib} amounts to a linear combination of $\chi_1^2$ variables whose relative weights
depend on estimated parameter and covariance values. Keep in
mind that $\boldsymbol{\Omega}_\mathbf{h}$
depends on $\lambda$  (see eq. \eqref{h_covariance}).

%\begin{proof} 
%
%It follows directly from the following version of
%the delta method (\cite{Serfling80}) for a real differentiable function  $g(\mathbf{a}(\lambda))$ (= $|\pi_{ij}(\lambda)|^2$) of the normal vectors $\mathbf{a}(\lambda)$ (guaranteed by Lemma \ref{lemma:ML1}) whereby 
%\begin{equation*}
%\sqrt{n_s}(g(\mathbf{a})_{n_s}-g(\mathbf{a}))\stackrel{d}{\rightarrow}\mathcal{N}(0,\mathbf{G}^T\;\mathbf{\Sigma} \;\mathbf{G})
%\end{equation*}
%where $\mathbf{G}=\nabla_\mathbf{a} g$ is the standard vector gradient of $g$ in equation (\ref{eq:quadraratio}) computed at $\mathbf{a}$. The proposition is obtained by straightforward computation of $\mathbf{G}$ recognizing $\mathbf{\Sigma}=\mathbf{\bar{\Omega}}$. The Slutsky's lemma allows using estimated quantities in lieu of the actual values.
%\end{proof}
\begin{flushright}\ensuremath{\Box}\end{flushright} 

%\noindent \textbf{Proof of Proposition \ref{prop:imhof}}
\begin{flushleft}
\textbf{Proof} 
\end{flushleft}

In view of the generalized \textit{delta} method version,
%(\cite{Serfling80}) from,
Theorem \ref{theo:gen_delta} whose conditions call for use of  $m=2$ under $H_0$
\eqref{eq:null_hyp} since both \eqref{eq:derivative_alpha}
and \eqref{eq:derivative_sigma_theo} become nullified.

First of all, note that taking derivative of \eqref{eq:derivative_alpha}
and \eqref{eq:derivative_sigma_theo} with respect to $\boldsymbol{\sigma}$ a second time does not alter the
$\mathbf{I}^c_{ij}\mathbf{a}$ dependence and so also produces null results. The same holds
when deriving \eqref{eq:derivative_sigma_theo} with respect to $\mathbf{a}$ since it is quadratic in
$\mathbf{I}^c_{ij}\mathbf{a}$. 

By contrast, the only nonzero surviving term is that of taking the derivative of \eqref{eq:derivative_alpha} with respect to
$\mathbf{a}$, which,
under $H_0$, reduces to
$$\frac{2}{\mathbf{h}^T\mathbf{I}_{i}^c\mathbf{\mathcal{S}}_d\mathbf{I}^c_{i}
\mathbf{h}}\mathbf{I}_{ij}^c
\mathbf{\mathcal{S}}_n\mathbf{I}_{ij}^c.$$

Therefore the Hessian in \eqref{eq:gendelta} only has an upper nonzero block corresponding to the derivative of \eqref{eq:derivative_alpha}
with respect to $\mathbf{h}$ so that one only needs to consider the distribution
of the latter to write
\begin{equation}
n_s(\hat{\mathbf{h}}^T\mathbf{I}^c_{j}\mathbf{\mathcal{S}}_d\mathbf{I}^c_{j}\hat
{ \mathbf{h}})(|\hat{\gamma}_{ij}|^2-|\gamma_{ij}|^2) \stackrel{d}{\rightarrow}
\mathbf{x}^T\mathbf{I}^c_{ij}\mathbf{\mathcal{S}}_n\mathbf{I}^c_{ij}\mathbf{x},
\end{equation}
using Theorem \ref{theo:gen_delta}
for $\mathbf{x}\stackrel{d}{\rightarrow}{\cal
N}(0,\boldsymbol{\Omega}_\mathbf{h})$.
The use of Slutsky's lemma concludes the first part of the proof by allowing the
use of estimated quantities.

Diagonalization of
$\mathbf{x}^T\mathbf{I}^c_{ij}\mathbf{\mathcal{S}}_n\mathbf{I}^c_{ij}\mathbf{x}$
is done via a transformation through the matrix $\mathbf{L}$ 
obtained from the Choleski decomposition of
$\boldsymbol{\Omega}_\mathbf{h}=\mathbf{L}\mathbf{L}^T$.
By making $\mathbf{x}=\mathbf{L}\mathbf{y}$, where
$\mathbf{L}=\mathbf{\mathcal{S}}^{1/2}_n\mathbf{I}^c_{ij}\mathbf{L}$,
 yields $\mathbf{x}^T\mathbf{I}^c_{ij}\mathbf{\mathcal{S}}_n \mathbf{I}^c_{ij}\mathbf{x}=
\mathbf{y}^T\mathbf{L}\mathbf{I}^c_{ij}\mathbf{\mathcal{S}}_n\mathbf{I}^c_{ij}\mathbf{L}\mathbf{y}
=\mathbf{y}^T\mathbf{D}\mathbf{y}$
so that the  elements of the vector $\mathbf{y}=(\mathbf{L}^T\mathbf{L})^{-1}\mathbf{L}^T\mathbf{x}$ are made mutually independent zero mean and of unit variance.
Now diagonalizing $\mathbf{D}=\mathbf{U}\boldsymbol{\Lambda}\mathbf{U}^T$ with
$\mathbf{U}\mathbf{U}^T=\mathbf{I}_{q\times q}$ produces
\begin{equation} \label{eq:quadform}
\mathbf{y}^T\mathbf{D}\mathbf{y}=\sum_{k=1}^q l_k \mathbf{y}^T \mathbf{u}_k\mathbf{u}^T_k\mathbf{y}=\sum_{k=1}^q l_k \zeta^2_k
\end{equation}
where $\mathbf{u}_k$ is the $k$-th column of $\mathbf{U}$. It is easy to show that the variables $\zeta_k=\mathbf{u}^T_k\mathbf{y}$ are mutually independent, normal zero mean and of unit variance so that $\zeta^2_k$ are $\chi^2_1$ random variables.

As $rank(\mathbf{X})=rank(\mathbf{X}^T)$ and 
$$rank(\mathbf{X}\mathbf{Y})\leq \min(rank(\mathbf{X}),rank(\mathbf{Y})),$$
it follows, after recalling explicit $\lambda$ dependence, that
\begin{eqnarray}
rank(\mathbf{D})&=&
rank(\mathbf{L}^T\mathbf{I}^c_{ij}\mathbf{\mathcal{S}}_n\mathbf{I}^c_
{ ij}\mathbf{L})\nonumber\\ 
&=&rank(\mathbf{L}\mathbf{L}^T\mathbf{I}^c_{ij}\mathbf{
\mathcal{S}}_n\mathbf{I}^c_{ij})\nonumber\\
&=&rank(\boldsymbol{\Omega}_\mathbf{h}(\lambda)\mathbf{I}^c_{ij}\mathbf{\mathcal
{ S}}_n\mathbf{I}^c_{ij})\nonumber\\
&=&rank(\mathbf{\mathcal{H}}(\lambda)
\mathbf{\mathcal{C}}(\lambda)\boldsymbol{\Omega}_{\boldsymbol{\alpha}}
\mathbf{\mathcal{C}}^T(\lambda)\mathbf{\mathcal{H}}^T(\lambda)\mathbf{I}^c_{ij}
\mathbf { \mathcal { S } } _n\mathbf { I } ^c_{ij}),
\end{eqnarray}
which is upper bounded by $rank(\mathbf{I}^c_{ij})=2$.
It is readily verified that, when $\lambda\in\{0,\pm0.5\}$,
$rank(\mathbf{\mathcal{C}}(\lambda))=1$ imposes the upper bound thus concluding the proof. As for the PDC case \cite{Baccala2013} if the model order $p=1$, though a detailed proof is more involved, one can show that the rank of $\mathbf{\mathcal{C}}^T(\lambda)\mathbf{\mathcal{H}}^T(\lambda)\mathbf{I}^c_{ij}$ equals 1 as it has only a single row that is not identically zero.
%In practice the number of degrees of freedom is adjusted to a single $\chi^2_\nu$ distribution with the appropriate number of degrees of freedom, see \cite{Takahashi2007} where fractional values of $\nu$ can be used to compute accurate thresholds under the Patnaik moment matching approximation.

%$\blacksquare$
%\end{proof}
\begin{flushright}\ensuremath{\Box}\end{flushright} 

\begin{center}
\textbf{Acknowledgements}
\end{center}
Work done under support from FAPESP (grant 2005/56464-9 -- CInAPCe Programme), CAPES Bioinformatics Graduate Programme (D.Y.T.),  FAPESP (grant 2008/08171-0 D.Y.T), Ci\^{e}ncia sem Fronteira (D.Y.T.), CNPq (grant 304404/2009-8 L.A.B. and grant 309381/2012 K.S.).
%\end{acknowledgements}

\bibliographystyle{plain}
%\bibpunct{(}{)}{;}{a}{,}{,}
%\bibliographystyle{unsrt}

\end{document}